\documentclass[12pt]{article}
\usepackage[margin=2.6cm]{geometry}
\usepackage[english]{babel} 
\usepackage[utf8]{inputenc}
\usepackage{indentfirst}
\usepackage{amsmath}
\usepackage{amssymb}
\usepackage{url}
\usepackage{graphicx}
\usepackage{braket}
\usepackage{slashed}
\usepackage{cancel}
\usepackage{ulem}
\usepackage[font=footnotesize,labelfont=bf]{caption}
\usepackage[usenames, dvipsnames]{xcolor}
\usepackage{booktabs}
\usepackage{dsfont}
\usepackage{pdflscape}
\usepackage{float}

\usepackage[colorlinks=true, urlcolor=MidnightBlue, linkcolor=MidnightBlue, 
citecolor=Mahogany, pdfborder={0 0 0}]{hyperref}
\usepackage[sort&compress,numbers]{natbib}
\usepackage{doi}
\usepackage[capitalise]{cleveref}

\newcommand{\DKpipi}{$D^+\to K^-\pi^+\pi^+$}
\newcommand{\DKpil}{$D^+\to K^-\pi^+ \ell^+\nu_\ell$}
\newcommand{\DKpie}{$D^+\to K^-\pi^+ e^+\nu_e$}
\newcommand{\DKpimu}{$D^+\to K^-\pi^+ \mu^+\nu_\mu$}
\newcommand{\DsKKpi}{$D_s^+\to K^+K^+\pi^-$}
\DeclareTextFontCommand{\comm}{\color{MidnightBlue}\em}

\newcommand{\nf}{na{\"i}ve factorization}

\title{\Large\bfseries{A combined study of hadronic \DKpipi\ and \DsKKpi\ decays
by means of the analysis of semileptonic \DKpil\ decays}}
\author{\normalsize{
        R.~Escribano{\color{Mahogany}\thanks{rescriba@ifae.es}}$^{\color{Mahogany}{\ a,b}}$, P.~Masjuan{\color{Mahogany}\thanks{masjuan@ifae.es}}$^{\color{Mahogany}{\ a,b}}$ 
        and
        Pablo Sanchez-Puertas{\color{Mahogany}\thanks{psanchez@ifae.es}}$^{\color{Mahogany}{\ b,c}}$
        }\vspace{0.2cm}\\
        {\small{$^{\color{Mahogany}{a}}$\textit{Grup de F{\'isica} Te{\`o}rica,
        Departament de F{\'i}sica,}}}\\
        {\small{\textit{Universitat Aut{\'o}noma de Barcelona, E-08193 Bellaterra (Barcelona), Spain}}}\\
        {\small{
        $^{\color{Mahogany}{b}}$\textit{Institut de F{\'i}sica d'Altes Energies (IFAE) and}}} \\
        {\small{\textit{
        Barcelona Institute of Science and Technology (BIST),}
        }}\\
        {\small{\textit{Campus UAB, E-08193 Bellaterra (Barcelona), Spain}}}\\
        {\small{
        $^{\color{Mahogany}{c}}$\textit{Departamento de F{\'i}sica At{\'o}mica, Molecular y Nuclear,
        }}}\\
        {\small{\textit{
        Universidad de Granada, E-18071, Spain}
        }}
}
\date{}

\begin{document}
\renewcommand{\abstractname}{\vspace{-\baselineskip}} 
\maketitle
\begin{abstract}
We perform a combined study of the two hadronic decays \DKpipi{} and \DsKKpi{} using a detailed analysis of the semileptonic decays \DKpil\ ($\ell=e, \mu$) thanks to the high-statistics 
dataset provided by the BESIII Collaboration.
We propose simple and suitable amplitude parametrizations of the studied reactions that shall be of interest to experimentalists for upcoming analyses.
These new parametrizations are based on the na{\"i}ve factorization hypothesis and the description of the resulting matrix elements in terms of well-known hadronic form factors, with special emphasis on the $K\pi$ scalar and vector cases. Such form factors account for two-body final state interactions which fulfill analyticity, unitarity and chiral symmetry constraints.
As a result of our study, we find that the $P$-wave contribution fits nicely within the na{\"i}ve-factorization approach, whereas the $S$-wave contribution requires complex Wilson coefficients that hint for possibly genuine three-body non-factorizable effects. Our hypothesis is further supported by the examination of \DsKKpi{} decays, where we achieve a description in overall good agreement with data.
\end{abstract}


\section{Introduction}
\label{intro}
In 2009, one of us presented a model for the decay \DKpipi\
where the weak interaction part of the reaction was described using the effective weak Hamiltonian in the na{\"i}ve factorization approach, while the two-body hadronic final state interactions were taken into account through the $K\pi$ scalar and vector form factors, fulfilling analyticity, unitarity and chiral symmetry constraints~\cite{Boito:2009qd}. However, due to the lack of precise data in semileptonic \DKpil{} decays ---a necessary ingredient in the na{\"i}ve factorization approach--- the model introduced two free parameters to describe the semileptonic form factor in terms of the scalar and vector $K\pi$ form factors, that were fixed from experimental \DKpipi{} branching ratios, preventing then a real prediction. Allowing for a global phase difference between the $S$ and $P$ waves, the Dalitz plot of the \DKpipi\ decay, the $K\pi$ invariant mass spectra, and the total branching ratio were well reproduced. Of course, lacking any input from semileptonic form factors, the model could not prove a real validation of the factorization hypothesis. Moreover, this motivates to generalize the (necessary) simplistic description in Ref.~\cite{Boito:2009qd} for such form factors, that is particularly relevant for the $S$ wave.

With the advent of new results for semileptonic decays by the BES-III Collaboration~\cite{Ablikim:2015mjo}, the whole model for the semileptonic form factor can be reviewed, and the performance of the factorization approach be tested, in contrast to Ref.~\cite{Boito:2009qd}. To that end, we carefully analyze semileptonic decays by employing simple yet well-motivated parametrizations fulfilling analyticity and unitarity constraints to fix the relevant hadronic matrix elements. These have their own interest for future experimental analysis. The corresponding matrix element, together with previously known form factors, is then used to describe the \DKpipi{} decay in the na{\"i}ve factorization approach.
As a result, we find that \nf{} describes well the $P$-wave contribution in \DKpipi{} decays for benchmark values of the Wilson coefficients, whereas the $S$-wave, that can also be effectively well described, forces us to incorporate complex Wilson coefficients. These are common anyway in $D$ decays~\cite{Cheng:2002wu,Qin:2013tje,Dedonder:2021dmb} and, in our opinion, point to non-factorizable corrections that might be attributed to effective genuine three-body effects. 
In this respect, it is worth emphasizing that our work is not meant to provide a precise and general description of these decays  (see Refs.~\cite{Magalhaes:2011sh,Guimaraes:2014kor,Magalhaes:2015fva,Nakamura:2015qga,Niecknig:2015ija} regarding 3-body unitarity effects missing here and Refs.~\cite{Diakonou:1989sf,Oller:2004xm} for previous works), but a first-order approximation that also allows to better understand the underlying fundamental QCD dynamics through the na{\"i}ve factorization approach. As a result, while our framework does not account for genuine three-body effects\footnote{The model does not account either for two-body $\pi^+\pi^+$ final state interactions, but these are non-resonant and presumably small, in such a way that na{\"i}ve factorization should encompass the most relevant two-body interactions.}, it allows for a simple parametrization fulfilling two-body unitarity and, not least, to connect \DKpipi{} decays to the isospin-related \DsKKpi{} ones. This actually allows to confront our hypothesis and results against \DsKKpi{} decays, improving and reinforcing our results.

The article is organized as follows: in \cref{sec:Formalism}, we outline the na{\"i}ve factorization approach applied to \DKpipi{} decays, recapitulating all the necessary form factors that enter the description; in \cref{sec:SL}, we review the semileptonic decays in detail, putting forward a parametrization that is used to extract the relevant form factors based on BES-III~\cite{Ablikim:2015mjo} results; in \cref{sec:HD}, we use the form factor from previous section to put forward a description for \DKpipi{} decays; in \cref{sec:HDs}, this parametrization is applied to the isospin related \DsKKpi{} decays. Conclusions are given in \cref{sec:conc}.

\section{Na{\"i}ve factorization in \DKpipi{} decays \label{sec:Formalism}}

\begin{figure}
    \centering
    \includegraphics[width=\textwidth]{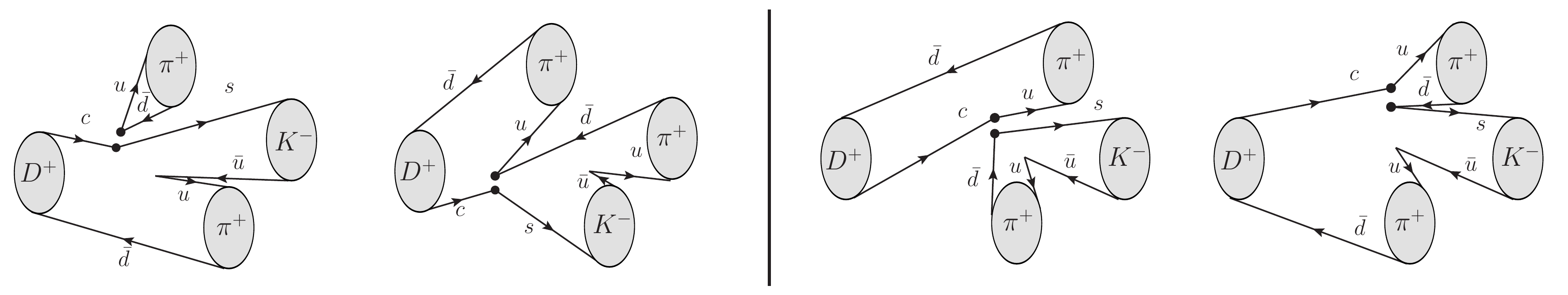}
    \caption{The $\mathcal{O}_1$ (left) and $\mathcal{O}_2$ (right) operator contributions to \DKpipi{} decays within na{\"i}ve factorization. For each operator there is a $N_c^0$- and $N_c^{-1}$-suppressed contribution, cf. left and right in each figure.}
    \label{fig:DKpipiFD}
\end{figure}

For \DKpipi decays, we closely follow Ref.~\cite{Boito:2009qd}. The effective weak interactions driving such decay follow from the Lagrangian at low energies
\begin{equation}
    \mathcal{L}_{\textrm{eff}} = -\frac{G_F}{\sqrt{2}}V_{ud}V_{cs}^* [C_1(\mu)\mathcal{O}_1 +C_2(\mu)\mathcal{O}_2] +\textrm{h.c.}\,, \qquad 
    \mathcal{O}_{1(2)} = 4[\bar{s}_{L}^{i}\gamma^{\mu} c_{L}^{i(j)}][\bar{u}_{L}^j\gamma^{\mu} d_{L}^{j(i)}] \,, 
\end{equation}
where $i,j$ are color indices, and the Wilson coefficients above differ from those at the electroweak scale due to renormalization~\cite{Buras:1994ij}. In the following, we employ the na{\"i}ve factorization hypothesis (see \cref{fig:DKpipiFD}), that implies the following decomposition for the process~\cite{Boito:2009qd}:
\begin{multline}\label{eq:DKpipiFact}
    i\mathcal{M} = -i\frac{G_F}{\sqrt{2}} V_{ud}V_{cs}^* \Big[ 
    a_1 \bra{K^-\pi^+_1} \bar{s}\gamma^{\mu}(1-\gamma^5)c \ket{D^+} 
        \bra{\pi^+_2} \bar{u}\gamma_{\mu}(1-\gamma^5)d \ket{0}  + \\
    a_2 \bra{K^-\pi^+_1} \bar{s}\gamma^{\mu}(1-\gamma^5)d \ket{0} 
        \bra{\pi^+_2} \bar{u}\gamma_{\mu}(1-\gamma^5)c \ket{D^+}
    \Big] +(\pi^+_1 \leftrightarrow \pi^+_2)\,,
\end{multline}
where na{\"i}ve factorization implies $a_1 = C_1 + C_2/N_c$, and $a_2 = C_2 + C_1/N_c$. While, ideally, these  coefficients should be universal, their scale and scheme dependence, together with potential non-factorizable corrections~\cite{Cheng:2002wu,Qin:2013tje}, render them somewhat phenomenological, with mild variations expected across different processes. Theoretically, Ref.~\cite{Buras:1994ij} obtained $a_1=1.31(19)$ and $a_2$ ranging between $-0.55(15)$ and $-0.60(22)$, depending on the chosen renormalization scheme. Alternatively, phenomenological processes can be used to determine them. For instance, an estimate coming from $D\to K\pi$ decays~\cite{Neubert:1991we,Buras:1994ij} obtains $a_1=1.2(1)$ and $a_2=-0.5(1)$, that can be considered as benchmark values. Different processes can be used to extract alternative determinations, whose agreement with previous numbers will allow to test the goodness of the na{\"i}ve-factorization approach and our understanding of QCD dynamics. Following \cref{eq:DKpipiFact}, factorization boils down the problem to the description of the following matrix elements in \cref{eq:DKpipiFact}: $\bra{\pi^+}\bar{u}\gamma^{\mu}(1-\gamma^5)d \ket{0} = i f_{\pi} p_{\pi}^{\mu}$ with $f_{\pi} =130.2(1.7)$~MeV~\cite{ParticleDataGroup:2022pth};  $\bra{K^-\pi^+} \bar{s}\gamma^{\mu}(1-\gamma^5)d \ket{0}$, that reduces to the well-known scalar and vector $K\pi$ form factors and that, following Ref.~\cite{Boito:2009qd}, we take from Refs.~\cite{Jamin:2006tj} and \cite{Boito:2010me}; $\bra{\pi^+} \bar{u}\gamma_{\mu}(1-\gamma^5)c \ket{D^+}$ is connected via isospin symmetry to $D^0\to \pi^- \ell^+ \nu$ decays; finally, the remaining matrix element, $\bra{K^-\pi^+_1} \bar{s}\gamma^{\mu}(1-\gamma^5)c \ket{D^+}$, corresponds to that appearing in semileptonic \DKpil{} decays. Indeed, a closer look reveals that all that is required for the current process is\footnote{In the last step, $i(m_s-m_c) \bra{K^-\pi^+} \bar{s}c \ket{D^+}=0$ has been used based on parity arguments.}
\begin{equation}
        i f_{\pi} p_{\pi_2}^{\mu} 
        \bra{K^-\pi^+_1} \bar{s}\gamma_{\mu}(1-\gamma^5)c \ket{D^+} 
        = f_{\pi}(m_c +m_s)\bra{K^-\pi^+_1}\bar{s}i\gamma^5c \ket{D^+}, \label{eq:DKpipiHard}
\end{equation}
that selects a single form factor among those appearing in semileptonic decays. It turns out that such a form factor produces a contribution to the semileptonic decays proportional to the lepton masses, that is irrelevant for \DKpie{} decays (see \cref{eq:WIgeneric,eq:F4trans,eq:I1,eq:I2,eq:I3,eq:I4,eq:I5,eq:I6,eq:I7,eq:I8,eq:I9}). 
Potentially, \DKpimu{} decays could probe such a form factor. At the moment, there is available data from FOCUS~\cite{Link:2002wg} 
and CLEO~\cite{Briere:2010zc}. Regarding FOCUS, the available statistics cannot discern a non-vanishing value for the form factor in \cref{eq:DKpipiHard}. Concerning CLEO, their results 
are controversial regarding the $q^2$-dependence. Thereby, some modelling is required. In the following we employ known relations due to Ward identities to suggest a plausible low-$q^2$ description based on existing results from semileptonic decays. To that end, in the following section we revise the model put forward in Ref.~\cite{Boito:2009qd} to describe the semileptonic matrix element, taking advantage of the precise results from BES-III not available at the time. This allows for a strict application of the \nf{} approach and a comprehensive evaluation of its performance.

\section{\DKpil{} decays \label{sec:SL}}

In this section, we address the semileptonic decay in detail, carefully reviewing the relevant form factors and paying special attention to the known restrictions that follow from Ward identities that, under reasonable assumptions, allow to extract the relevant form factor entering hadronic decays. 
Our phenomenological description generalizes that in Ref.~\cite{Boito:2009qd} by incorporating free parameters previously identified with those appearing in $K\pi$ form factors  ---a necessary assumption back then in the absence of data that can be relaxed now by using the recent results from BES-III~\cite{Ablikim:2015mjo}.

\subsection{General definitions}

The matrix element for semileptonic decays is given as~\cite{Lee:1992ih}\footnote{Note our $\epsilon^{0123}=1$ convention, leading to opposite signs compared to Ref.~\cite{Lee:1992ih} wherever the antisymmetric tensor appears (the sign can be inferred from $L^{\mu\nu}$). We also employ $\epsilon^{\mu k p q} \equiv \epsilon^{\mu\nu\alpha\beta}k_{\nu}p_{\alpha}q_{\beta}$.}
\begin{equation}\label{eq:Dl4M}
 \mathcal{M} = -\frac{G_F}{\sqrt{2}} V_{cs}^* \bra{\pi^+ K^-} \bar{s}\gamma^{\mu}(1-\gamma^5)c \ket{D^+} 
                 [\bar{u}_{\nu}\gamma_{\mu}(1-\gamma^5)v_{\ell}] 
              \rightarrow |\mathcal{M}|^2 = 4G_F^2|V_{cs}|^2 H^{\mu\nu}L_{\mu\nu}\,,
\end{equation}
where we used ($p_{\ell\nu} = p_{\ell}+p_{\nu}$ and $\bar{p}_{\ell\nu} = p_{\ell}-p_{\nu}$) and
\begin{align}
    H^{\mu\nu} ={}& \bra{\pi^+ K^-} \bar{s}\gamma^{\mu}(1-\gamma^5)c \ket{D^+}
                 \bra{\pi^+ K^-} \bar{s}\gamma^{\nu}(1-\gamma^5)c \ket{D^+}^{\dagger}, \\
    L^{\mu\nu} ={}& \frac{1}{2}\left[ 
                  p_{\ell\nu}^{\mu}p_{\ell\nu}^{\nu} -\bar{p}_{\ell\nu}^{\mu}\bar{p}_{\ell\nu}^{\nu}
                  -(s_{\ell\nu} -m_{\ell}^2 -m_{\nu}^2)g^{\mu\nu} +i\epsilon^{p_{\ell\nu}\mu\bar{p}_{\ell\nu}\nu}
                 \right].
\end{align}
As such, the central quantity is the matrix element in \cref{eq:DKpipiFact} which, using the variables $p= p_K +p_{\pi}$, $\bar{p}= p_K -p_{\pi}$, and $q=p_D -p$, can be expressed as~\cite{Lee:1992ih,Bajc:1997nx}
\begin{align}
    \bra{K^-\pi^+} \bar{s}\gamma^{\mu}(1-\gamma^5)c\ket{D^+} ={}& 
     iw_+p^{\mu} +iw_-\bar{p}^{\mu} +ir q^{\mu} -h\epsilon^{\mu q p \bar{p}}\nonumber\\
    ={}& 
     iw_+(p^{\mu}\! -q^{\mu}\frac{p\!\cdot\! q}{q^2}) 
   +iw_-(\bar{p}^{\mu}\! -q^{\mu}\frac{\bar{p}\!\cdot\! q}{q^2}) 
   +\frac{i\tilde{r}}{q^2} q^{\mu}  -h\epsilon^{\mu q p \bar{p}}\,, \label{eq:DKpi}
\end{align}
where the four form factors have an implicit dependence on $q^2$, $p^2$, and $\bar{p}\cdot q$. Note that corresponding quantities in $D^-$ decays are related via appropriate $CP$ transformations that amount to flip signs for the antisymmetric tensor. In addition, the Ward identities (\textit{i.e.}, \cref{eq:DKpipiHard} and finiteness at $q^2=0$) imply
\begin{equation}\label{eq:WIgeneric}
    \tilde{r} = -(m_c+m_s) \bra{K^-\pi^+} \bar{s}i\gamma^5c \ket{D^+}, \qquad 
    \lim_{q^2\to0} [(p\cdot q)w_+ +(\bar{p}\cdot q)w_- -\tilde{r}] = 0\,.
\end{equation}
In particular, their dependence on $\bar{p}\cdot q \sim \cos\theta_{K\pi}$ (see \cref{app:DSLdecays}) means that the relation should be fulfilled for each partial wave (see also Ref.~\cite{Lee:1992ih}), a property that we will employ when constructing the form factors. Note that the appearance of $\bar{p}$ in the tensor structure accompanying $w_-$ and $h$ requires partial-wave contributions with $\ell \geq 1$.
To make contact with experiment, it is customary to employ the following form factors~\cite{Lee:1992ih,Bajc:1997nx}\footnote{Note in this respect that for the kinematic variables chosen for the semileptonic decay, $X^2 = (p\cdot q)^2-p^2q^2$, while $[(p\cdot q)(\bar{p}\cdot q) -q^2(p\cdot\bar{p})] = X(z\beta_{K\pi}\cos\theta_{K\pi} + X\Delta_{K\pi})$, reproducing the result in Ref.~\cite{Bajc:1997nx}. However, we keep it general in order to use it in \DKpipi{} decays.} (see definitions in \cref{app:DSLdecays})
\begin{align}
    F_1 ={}& \frac{1}{X}\left(  X^2w_+  +[(p\cdot q)(\bar{p}\cdot q) 
                             -q^2(p\cdot\bar{p})]w_-\right)\,,\label{eq:F1trans}\\[1ex]
    F_2 ={}& \beta_{K\pi}\sqrt{s_{K\pi}s_{\ell\nu}} w_-\,,\label{eq:F2trans}\\[1ex]
    F_3 ={}& \beta_{K\pi} X \sqrt{s_{K\pi}s_{\ell\nu}} h\,,\label{eq:F3trans}\\[1ex]
    F_4 ={}& \tilde{r}\,.\label{eq:F4trans}
\end{align}
where $F_i\equiv F_i(q^2,p^2,\bar{p}\cdot q)$. $F_4$ was not defined in Ref.~\cite{Lee:1992ih} and is only relevant for finite lepton masses, so that results in \cref{app:DSLdecays} might be of some interest. Further, \cref{eq:WIgeneric} implies for these form factors that 
\begin{equation}\label{eq:F1F4WI}
        \lim_{q^2\to 0}[F_1(q^2,p^2,\bar{p}\cdot q) -F_4(q^2,p^2,\bar{p}\cdot q)] = 0\,, 
\end{equation}
that relates again the normalization at $q^2=0$ that, as mentioned, must be fulfilled for each partial wave. 
This is as far as can be reached in a model-independent way and we refer to \cref{app:DSLdecays} for the differential decay width expressed in terms of the previous form factors. In the following section, we {\textcolor{red}{revise}} the model that was used in Ref.~\cite{Boito:2009qd} to parametrize $F_4$ that, in essence, assumes the $K\pi$ spectra to be dominated by intermediate resonances with roles parallel to those in $\bra{K^-\pi^+} \bar{s}\gamma^{\mu} d \ket{0}$ form factors. In doing so, we employ a more flexible description compared to that in Ref.~\cite{Boito:2009qd}, that will prove convenient to make contact with the standard phenomenological analysis.

\subsection{Resonance model}\label{sec:model}

With the lack of precise data for semileptonic decays, Ref.~\cite{Boito:2009qd} assumed a model for the $F_4$ form factor saturated by the lightest $K_{(0)}^*$ resonances. Assuming a similar model for the $K\pi$ scalar and vector form factor allowed them to relate the $p^2$ dependence of the previous form factors to that of the $K\pi$ scalar and vector form factors. While this was a necessary assumption back then, the current available data for semileptonic decays from BES-III~\cite{Ablikim:2015mjo} allows to relax this assumption and to provide a more flexible and realistic model based on analyticity and unitarity, that might be of interest for future experimental analysis. In particular, in the following we assume that the $S$- and $P$-waves contributions share the same phase as the scalar and vector form factor (that holds below threshold due to Watson's theorem) but have, in general, a different $q^2$-dependence compared to the $K\pi$ scalar and vector form factors.

\subsubsection{Scalar contributions}

In the original work from Ref.~\cite{Boito:2009qd}, the scalar contribution was assumed to be dominated by the $K_0^*(1430)$~\cite{ParticleDataGroup:2022pth} resonance, whose peak dominates the $K\pi$ scalar form factor, $F_0^{K\pi}(s)$, at intermediate energies. Under the assumption that such a resonance is a quasi-stable (e.g. narrow) state, the $D^+\to \bar{K}_0^* \ell^+\nu$ decay can be described via its matrix element ($p$ is the momentum associated to the $\bar{K}_0^*$)
\begin{equation}
    \bra{\bar{K}^*_0} \bar{s}\gamma^{\mu}(1-\gamma^5)c\ket{D^+} = 
           i\left[ w_+^{\bar{K}_0^*}(q^2)\left(p_{\bar{K}_0^*}^{\mu} - \frac{q\cdot p_{\bar{K}_0^*}}{q^2}q^{\mu}\right) + q^{\mu}\frac{\tilde{r}^{\bar{K}_0^*}(q^2)}{q^2} \right],
\end{equation}
where once more 
\begin{equation}\label{eq:WIscalar}
    \tilde{r}^{\bar{K}_0^*}(q^2) = -(m_c+m_s)\bra{\bar{K}_0^*} \bar{s}i\gamma^5 c \ket{D^+}, \qquad
    \lim_{q^2\to 0} \left[(q\cdot p) w_+^{\bar{K}_0^*}(q^2)- \tilde{r}^{\bar{K}_0^*}(q^2)\right] =0\, .
\end{equation}
Finally, the $q^2$ dependence is reduced, as usual, to the closest charmonium resonance. Its subsequent $\bar{K}_0^*\to K^-\pi^+$ decay merely adds the resonance structure, meaning that the full amplitude is given as $ \braket{K^-\pi^+ |\bar{K}_0^*} P_{\bar{K}_0^*} \bra{\bar{K}^*_0} \bar{s}\gamma^{\mu}(1-\gamma^5)c\ket{D^+}$, with $P_{\bar{K}_0^*}$ the scalar propagator. Were this fully dominated by the $K_0^*$ resonance both for the semileptonic and $K\pi$ scalar form factors, then $ \braket{K^-\pi^+ |\bar{K}_0^*} P_{K_0^*} \to \chi_{\bar{K}_0^*} F_0^{K\pi}$, where $\chi_{\bar{K}_0^*} =(m_K^2 -m_{\pi}^2)/(m_s -m_d)/\bra{\bar{K}_0^*} \bar{s}d \ket{0}$~\cite{Boito:2009qd}. However, the different interplay of scalar resonances shall in general differ, yet their phase shift below inelasticities must agree by Watson's theorem. We reflect this by shifting $F_{0}^{K\pi} \to F_{0}^{D_{\ell 4}}$, that allows for the following {\textit{ansatz}} for the $S$-wave contribution
\begin{align}
    w_+(q^2,p^2,\bar{p}\cdot q) ={}& X^{-1}F_1^{\bar{K}_0^*}(q^2,p^2,\bar{p}\cdot q)= 2\chi_S^{\textrm{eff}}  F_{0}^{D_{\ell 4}}(p^2) \left( 1-q^2/m^2_{D_{s1}} \right)^{-1}, \label{eq:F1Scalar}\\[1ex]
    \tilde{r}(q^2,p^2,\bar{p}\cdot q) ={}& F_4^{\bar{K}_0^*}(q^2,p^2,\bar{p}\cdot q)= \chi_S^{\textrm{eff}}(m_D^2-p^2)  F_{0}^{D_{\ell 4}}(p^2) \left( 1-q^2/m^2_{D_{s}} \right)^{-1}.\label{eq:F4Scalar} 
\end{align}
The parametrization in \cref{eq:F4Scalar} has been chosen to fulfill \cref{eq:WIgeneric,eq:WIscalar}, and to include the closest pole with appropriate quantum numbers.
Regarding the parametrization used in BES~III~\cite{Ablikim:2015mjo}, we identify $2\chi_S^{\textrm{eff}}F_0^{D_{\ell4}}(p^2) = \mathcal{A}_S(p^2)$ (see Eq.~(20) from Ref.~\cite{Ablikim:2015mjo}).

In contrast to Ref.~\cite{Boito:2009qd}, in order to parametrize $F_{0}^{D_{\ell 4}}(p^2)$, we follow the approach in Refs.~\cite{Bernard:2006gy,Bernard:2009zm,Bernard:2013jxa}. This uses an Omn{\`e}s representation subtracted at $p^2=0$ and the Callan-Treiman point $\Delta_{K\pi} = m_K^2 -m_{\pi}^2 $, 
\begin{align}\label{eq:scalarmodel}
    F_{0}^{D_{\ell 4}}(s) &= \operatorname{exp}\left[
    \frac{s[\ln C_{D_{\ell4}} +G_0(s)]}{\Delta_{K\pi}}
    \right],  \\[1ex]
    G_0(s) &=  
    \frac{\Delta_{K\pi}(s-\Delta_{K\pi})}{\pi}
    \int_{s_{th}}^{\infty} d\eta \frac{\delta_0^{1/2}(\eta)}{\eta(\eta -\Delta_{K\pi})(\eta -s)}\,,
\end{align}
with $\delta_0^{1/2}$ the scalar $I=1/2$ $K\pi$ phase shift, that preserves the constraints provided by unitarity and analyticity below higher inelasticities. The subtraction constant, $\ln C_{D_{\ell4}}$, encapsulates high-energy effects that need not be the same as in the $K\pi$ scalar form factor case, thus requiring data on semileptonic decays to fix it. For the phase shift, we take that in Ref.~\cite{Jamin:2006tj} below $\Lambda=1.67$~GeV, where $\delta_0^{1/2}(\Lambda) =\pi$; above, we take a constant phase $\delta_0^{1/2} =\pi$ following Ref.~\cite{Bernard:2006gy,Bernard:2009zm,Bernard:2013jxa}.
This model allows for a relatively simple and flexible parametrization, that improves the one used by the BES-III Collaboration by incorporating appropriate analyticity and unitarity constraints (up to higher-threshold inelasticities). As such, it might be useful in future experimental analyses.

\subsubsection{Vector contributions}

The next relevant wave is the $P$-wave, where the narrow $\bar{K}^*$ resonance plays a prominent role both in the $K\pi$ vector form factor and semileptonic decays. Again, assuming them to be narrow states, the $D^+\to \bar{K}^* \ell^+\nu$ decay can be described via the corresponding matrix element\footnote{Different parametrizations appear in Refs.~\cite{Wirbel:1985ji,Bernard:1991bz,Bowler:1994zr,Richman:1995wm,Melikhov:2000yu,Bajc:1995km,Khodjamirian:2020btr}; the connection reads, up to overall signs,
\begin{equation}
    A = -\frac{2V}{m_{\bar{K}^*}(m_D +m_{\bar{K}^*})}\,,   \quad
    B = \frac{m_D +m_{\bar{K}^*}}{m_{\bar{K}^*}}A_1\,,  \quad
    C = -\frac{2A_2}{m_{\bar{K}^*}(m_D +m_{\bar{K}^*})} \,, \quad
    \tilde{D}= 2A_0\,.
\end{equation}}
\begin{equation} \label{eq:DK*FF}
    \bra{\bar{K}^{*}} \bar{s}\gamma^{\mu}(1-\gamma^5)c\ket{D^+}
     \!= \! \Big( \!A\epsilon^{\mu\nu qp} 
           - i\Big[ B(g^{\mu\nu} \! -\frac{q^{\mu}q^{\nu}}{q^2}) 
           + Cq^{\nu}( p^{\mu} -\frac{q\!\cdot\! p}{q^2}q^{\mu}) 
            + \frac{\tilde{D}}{q^2}q^{\mu}q^{\nu}\Big] \Big)m_{\bar{K}^*}\varepsilon_{\nu}\,, 
\end{equation}
where $m_{\bar{K}^*}$ has been used for later convenience. In addition, the Ward identity implies 
\begin{align}
    m_{\bar{K}^*}\tilde{D}(q^2)(q\cdot \varepsilon) = (m_c+m_s)\bra{K^*} \bar{s}i\gamma^5 c \ket{D^+}, \\ 
    \lim_{q^2\to0} \left[ B(q^2) +(q\cdot p) C(q^2) - \!\tilde{D}(q^2) \right]\! =0\,.
\end{align}
Again, the $q^2$-dependence can be saturated via the appropriate charmonium resonances. Then, along the lines in Ref.~\cite{Boito:2009qd}, the subsequent $K^-\pi^+$ decay would closely resemble the vector $K\pi$ form factor if both cases were fully dominated by the $K^*$. 
Still, as for the scalar case, these will generally differ ---even if the phase shift below inelasticities should be the same. Therefore, we replace once more $F_+^{K\pi}(p^2)\to F_+^{D_{\ell4}}(p^2)$ (for a single resonance contribution $\chi_{\bar{K}^*} = f_{\bar{K}^*}^{-1}$~\cite{Boito:2009qd}), obtaining
\begin{align}
    F_1(q^2,p^2,\bar{p}\cdot q) =&{} 
       -\!\chi_{\bar{K}^*}F_+^{D_{\ell4}}(p^2)  \beta_{K\pi}\cos\theta_{K\pi}\left[X^2C(0)+ (q\cdot p) B(0) \right] \left[1- q^2/m_{D_{s1}}^2\right]^{-1}, \label{eq:F1V}\\[1ex]
    F_2(q^2,p^2,\bar{p}\cdot q) =&{} 
       -\!\chi_{\bar{K}^*}F_+^{D_{\ell4}}(p^2)  \beta_{K\pi}\sqrt{s_{K\pi} s_{\ell\nu}}B(0) \left[1- q^2/m_{D_{s1}}^2\right]^{-1}, \label{eq:F2V}\\[1ex]
    F_3(q^2,p^2,\bar{p}\cdot q) =&{} 
       \!-\chi_{\bar{K}^*}F_+^{D_{\ell4}}(p^2)  \beta_{K\pi} X\sqrt{s_{K\pi} s_{\ell\nu}}A(0) \left[1- q^2/m_{D_{s}^*}^2\right]^{-1}, \label{eq:F3V}\\[1ex]
    F_4(q^2,p^2,\bar{p}\cdot q) =&{} 
       \!-\chi_{\bar{K}^*}F_+^{D_{\ell4}}(p^2) \frac{N(p^2)}{2} \tilde{D}(q^2,p^2) \nonumber\\ 
             =&{} -\chi_{\bar{K}^*}F_+^{D_{\ell4}}(p^2) \frac{N(p^2)}{2} \left[B(0) +\frac{m_D^2 -p^2}{2}C(0)\right] \left[1- q^2/m_{D_{s}}^2\right]^{-1}, \label{eq:F4V}
\end{align}
where $N(p^2)/2= [p^2(\bar{p}\cdot q) -(p\cdot \bar{p})(p\cdot q)]/p^2$ is a variable defined in Ref.~\cite{Boito:2009qd} that reduces to $X\beta_{K\pi}\cos\theta_{K\pi}$ in semileptonic decays, and the last form factor is chosen to fulfill \cref{eq:F1F4WI} and saturated with the closest resonance{.\footnote{Further, the $q^2$ dependence is chosen to vanish asymptotically since, otherwise, the $\ell \nu\to DK^*$ amplitude would grow indefinitely. Moreover, this dependence matches the pole-dominance behavior that would be assigned for a pseudoscalar form factor.} Note also the different $p^2$-dependence with respect to the one in Ref.~\cite{Boito:2009qd}. 
The connection to the {\textit{ansatz}} employed by the BES-III Collaboration~\cite{Ablikim:2015mjo} can be easily obtained accounting that  $2\alpha\sqrt{2} m^{-1}\mathcal{A}(m) = -g_{\bar{K}^*K\pi}\beta_{K\pi}P_{\bar{K}^*}(m^2)$, with $P_{\bar{K}^*}(s)$ the standard propagator.
Once again, to obtain a description fulfilling appropriate analyticity and unitarity constraints below higher inelasticities, we take 
\begin{equation}
   F_+^{D_{\ell4}}(s) = \operatorname{exp}\left[ \lambda_1\frac{s}{m_{\pi}^2} + G_+(s)\right], \quad 
   G_+(s) = \frac{s^2}{\pi} \int_{s_{th}}^{\infty} d\eta \frac{\delta_1^{1/2}(\eta)}{\eta^2(\eta -s)}\,,
\end{equation}
with $\delta_1^{1/2}$ the $P$-wave $I=1/2$ $K\pi$ phase shift.
The input for the phase shift is taken from the result in Ref.~\cite{Escribano:2014joa} with a single vector resonance and with a single subtraction constant. To match their results, we choose an upper cutoff $s=4~\textrm{GeV}^2$ and $\lambda_1=0.025$, but such parameter could be fitted from the experiment, providing then an useful parametrization for experimentalists. Further details are given in \cref{sec:vectorFF}.

\subsection{Extraction of the parameters from BES-III}

Since there is no available data from experiment (that should be also unfolded), we {have to restrict ourselves to fit our model to the scalar and vector form factors extracted by the BES-III Collaboration. 
Still, we emphasize that having such data available would allow for a more reliable estimate of our parameters {and a more robust analysis for the $S$-wave compared to Ref.~\cite{Ablikim:2015mjo}.
Regarding the free parameters for the scalar part (cf. \cref{eq:scalarmodel}), we fit $2\chi_{S}^{\textrm{eff}} F_0^{D_{\ell4}}(s)$ to pseudodata from the $A_S(s)$ form factor from BES-III, obtaining 
\begin{equation}
\chi_{S}^{\textrm{eff}}=2.13(16)~\textrm{GeV}^{-1}\,, \quad 
\ln C_{D_{\ell4}} = 0.152(11)\,, 
\end{equation}
with a correlation of $-0.27$. We show our results in \cref{fig:ScalarFFsl}. 
\begin{figure}
    \centering
    \includegraphics[width=0.495\textwidth]{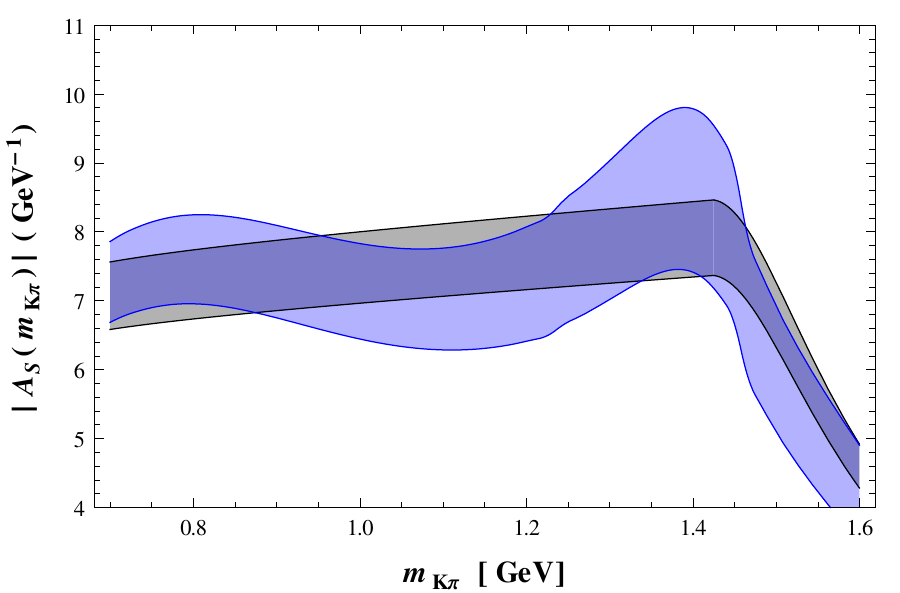}
    \includegraphics[width=0.495\textwidth]{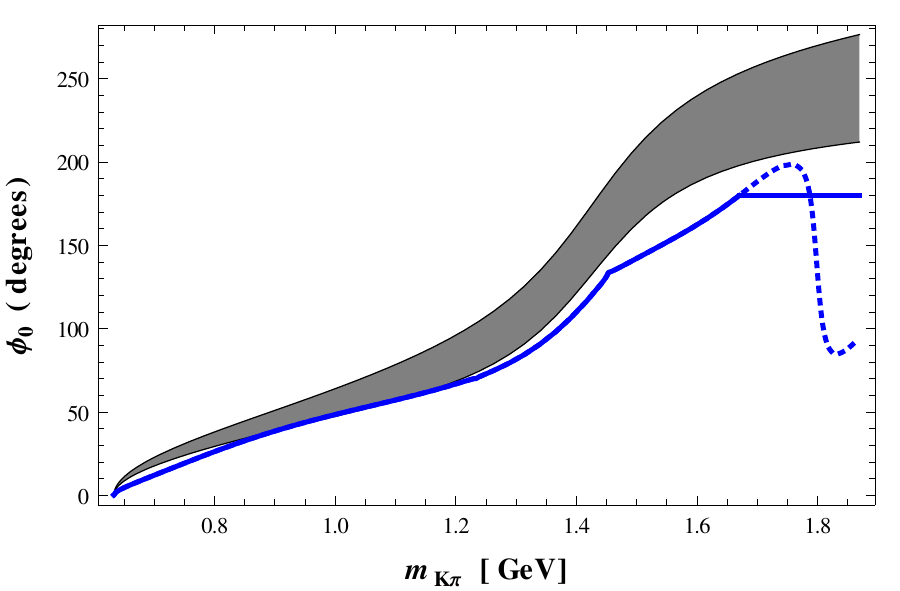}
    \caption{Modulus (left) and phase (right) of the scalar form factor. The gray band stands for BES-III results~\cite{Ablikim:2015mjo}, while the blue band represents our model. The dotted line in the phase plot represents the original input from \cite{Jamin:2006tj}. We neglect errors from the phase shift that are subleading as compared to BES-III uncertainties on the modulus. 
    }
    \label{fig:ScalarFFsl}
\end{figure}
Here, it is worth emphasizing two aspects. First, regarding the right-panel, the $\delta_0^{1/2}$ $K\pi$ phase-shift has been taken from the sophisticated analysis in Ref.~\cite{Jamin:2000wn,Jamin:2006tj}, in contrast with Ref.~\cite{Ablikim:2015mjo} that uses an effective-range expansion~\cite{Bethe:1949yr}. Second, with regard to the $m_{K\pi}$-behavior depicted in the left-panel, it is important to note that the two models must necessarily differ by construction. Indeed, Ref.~\cite{Ablikim:2015mjo} assumes explicitly a linear dependence below the $\bar{K}_0^*(1430)$, that contrasts with the behavior dictated by analyticity and unitarity, that is fulfilled by the Omn{\`e}s-like solution but not in their simplified model, and is ultimately responsible for the differences. The crucial point here is whether our predicted differential spectra (see \cref{fig:SLdiffSpectra}) compares well to data~\cite{Ablikim:2015mjo}. Whereas the data is well described by the model in Ref.~\cite{Ablikim:2015mjo}, the data precision is well below the one that would be inferred from their mdel away from the $\bar{K}^*(892)$ resonance.
Note that, in the case of $F_0^{K\pi}(s)$, the chosen parametrization would require $\ln C_{D\ell4}=0.206(9)$~\cite{Bernard:2013jxa}, based on a combined analysis from $\tau\to K\pi\nu$ and $K_{\ell3}$ decays. This is not incosistent, but shows that the necessary assumption adopted back in Ref.~\cite{Boito:2009qd} holds only approximately. Concerning the vector part, we fit the differential decay width distributions obtained from pseudodata from BES-III parametrization with vector contributions only. This way we obtain the parameters
\begin{equation}
    \chi_A^{\textrm{eff}} =-3.35(16)~\textrm{GeV}^{-3}\,, \quad 
    \chi_B^{\textrm{eff}} 
    =8.44(23)~\textrm{GeV}^{-1}\,, \quad
    \chi_C^{\textrm{eff}} 
    = -1.64(12)~\textrm{GeV}^{-3}\,,
\end{equation}
where $\chi_X^{\textrm{eff}} \equiv \chi_{\bar{K}^*}X(0)$ in \cref{eq:F1V,eq:F2V,eq:F3V,eq:F4V}. The error to describe the semileptonic decay is fully dominated by that of $\chi_B^{\textrm{eff}} $. The correlation for $\chi_B^{\textrm{eff}}$ and $\chi_C^{\textrm{eff}}$, that enters the $F_4$ form factor, reads -$0.35$.
\begin{figure}
    \centering
    \includegraphics[width=0.8\textwidth]{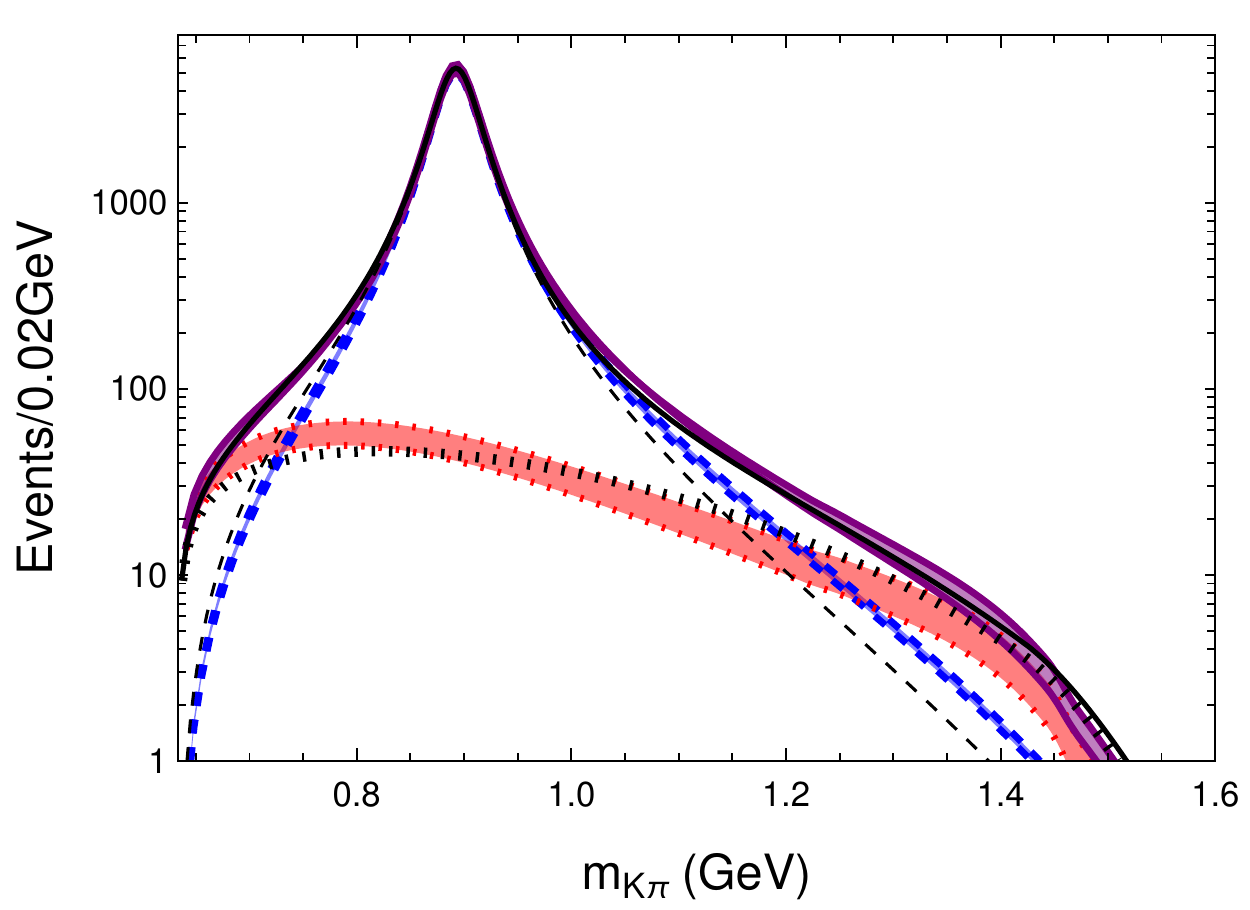}
    \caption{The differential spectra normalized to BES-III events. The red band is our scalar contribution, the blue one is the vector part and the purple band is their combination. The black dotted, dashed, and full lines stands for the scalar, vector and full BES-III model, respectively.
    }
    \label{fig:SLdiffSpectra}
\end{figure}
In \cref{fig:SLdiffSpectra}, we show our description for the differential $m_{K\pi}$ spectrum compared to the central values of BES-III, showing nice agreement. We observe that an overall sign cannot be extracted from experiment; to do so, we make use of quark models~\cite{Melikhov:2000yu,Albertus:2005vd}, that allow to choose a sign that is consistent among the different matrix elements here considered. These imply a positive sign for $\chi_B^{\textrm{eff}}$. That this gives the correct interference pattern in the hadronic decays below {can be considered as a positive test of the na{\"i}ve factorization hypothesis

\section{Case I: \DKpipi{} decays\label{sec:HD}}

The analysis presented in the preceding section, combined with the pertinent form factors introduced in \cref{sec:Formalism}, provides all the necessary inputs for the application of the na{\"i}ve factorization approach, an accomplishment that was not possible in Ref.~\cite{Boito:2009qd}. That allows the possibility to strictly test its performance. The relevant matrix element are summarized in the following,
\begin{align}
    \bra{K^-\pi^+} \bar{s}\gamma^{\mu}(1\!-\gamma^5)d \ket{0} \! &= \!\left( \bar{p}_{K\pi}^{\mu} \! - \frac{\Delta_{K\pi}}{p_{K\pi}^2} p_{K\pi}^{\mu} \right)\! F_+^{K\pi}(p_{K\pi}^2) + \frac{\Delta_{K\pi}}{p_{K\pi}^2} p_{K\pi}^{\mu} F_0^{K\pi}(p_{K\pi}^2)\,, \\[1ex]
    \bra{\pi^+} \bar{u}\gamma_{\mu}(1\!-\gamma^5)c \ket{D^+} \! &= \!\left( p_{D\pi}^{\mu} -\frac{\Delta_{D\pi}}{\bar{p}_{D\pi}^2} \bar{p}_{D\pi}^{\mu}  \right)F_+^{D\pi}(\bar{p}_{D\pi}^2) + \frac{\Delta_{D\pi}}{\bar{p}_{D\pi}^2} \bar{p}_{D\pi}^{\mu} F_0^{D\pi}(\bar{p}_{D\pi}^2)\,,\\[1ex]
    \hspace{-.15cm}i f_{\pi} p_{\pi_2}^{\mu} 
        \bra{K^-\pi^+_1} \bar{s}\gamma_{\mu}(1\!-\gamma^5)c \ket{D^+} \! &= \! - \! f_{\pi}F_4 = 
        -f_{\pi} \Big[\chi_S^{\textrm{eff}}(m_D^2 -p_{K\pi}^2)F_{0}^{D_{\ell4}}(p_{K\pi}^2) \nonumber \\ & 
              \hspace{-1.5cm}  -\frac{1}{2}N(p_{K\pi}^2)F_{+}^{D_{\ell4}}(p_{K\pi}^2) \Big( \chi_B^{\textrm{eff}} +\frac{m_D^2 -p_{K\pi}^2}{2}\chi_C^{\textrm{eff}} \Big)  \Big]\frac{1}{1- m_{\pi}^2/m_{D_s}^2}\,,
\end{align}
where $p_{AB}^{\mu} = p_A^{\mu} +p_B^{\mu}$, $\bar{p}_{AB}^{\mu} = p_A^{\mu} -p_B^{\mu}$ and $\Delta_{AB}=m_A^2 -m_B^2$. 
The form factors $F_{+,0}^{D_{\ell4}}$ are those from the previous section. 
Concerning $F_{0}^{K\pi}(s)$, we use that from Ref.~\cite{Jamin:2006tj}, while for $F_{+}^{K\pi}(s)$ we take that from Ref.~\cite{Escribano:2014joa}. For the $D^+\to\pi^+$ transition, we use isospin symmetry that relates it to that in $D^0\to \pi^- \ell^+ \nu$ decays, that is parametrized as
\begin{equation}
    F_{+(0)}^{D\pi}(s) = \frac{F_{+(0)}^{D\pi}(0)}{1- s/m^2_{D_{(0)}^{*0}}}\,,
    \quad F_{0}^{D\pi}(0) = F_{+}^{D\pi}(0) = 0.612(35)~\textrm{\cite{Lubicz:2017syv}}\,.
\end{equation}
The final result for the amplitude and differential decay width can be easily obtained by applying \cref{eq:DKpipiFact} and is given in \cref{app:DKpipi}. With the necessary expressions at hand, we proceed to precisely fix the Wilson coefficients as inferred by this process, which comparison to benchmark values will provide valuable insight about \nf{} and our understanding of the QCD dynamics. 
To do so, it is instructive to work out the $P$-wave contribution first and the $S$-wave subsequently.

\subsection{P-wave contribution}

The $P$-wave contribution, with a branching ratio of $1.06(12)\%$~\cite{ParticleDataGroup:2022pth}, is fully dominated by the $\bar{K}^{*0}(892)$ resonance. As such, it essentially corresponds to that of a quasi-two-body $D^+\to\bar{K}^{*0}(892)\pi^+$ decay and should be relatively free of genuine three-body problems (that amount to non-factorizable corrections). Consequently, it represents a theoretically clean observable compared to the $S$-wave, which discussion is relegated to the following section. 
Before extracting the Wilson coefficients, it is instructive to estimate first the results for benchmark $a_{1,2}$ values. 
Taking the aforementioned theory estimate, $a_1=1.31(19)$, $a_2= -0.55(30)$, we obtain $\operatorname{BR}=(0.24^{+1.22}_{-0.30}) \%$, perfectly consistent with the experiment, albeit with large uncertainties. The phenomenological estimate ($a_1=1.2(1)$, $a_2=-0.5(1)$) points towards a lower value, $\operatorname{BR}=(0.19^{+0.42}_{-0.22})\%$, but once again with large uncertainties. Overall, it seems that the $P$-wave BR is consistent with benchmark estimates. However, its specific value proves to be highly sensitive to the Wilson coefficients, making it challenging to obtain a precise and reliable prediction in this manner. As such, after this detour, we finally proceed to extract the Wilson coefficients by demanding the experimental $P$-wave BR be fulfilled, that requires (errors are omitted here) 
\begin{equation}
(2.87 a_1^2 + 10.55 a_2^2 + 10.98 a_1 a_2)\% = 1.06\%\, .
\end{equation}
This admits several possibilities for $\{a_1,a_2\}$, such as $\{1.56,-0.5\}$, $\{1.2,-0.21\}$, or $\{1.31$, $-0.37\}$, all falling within the ballpark estimates. This represents a positive indication of the effectiveness of the na\"ive factorization approach. It must be emphasized that the relative sign between the $a_{1,2}$ contributions constitutes a prediction within this framework. Would this sign be opposite, one would require $a_1\leq 0.61$ and $|a_2| \leq 0.32$ (with the equal sign achieved for either $a_{1,2}=0$), that significantly deviates from the benchmark values for $a_1$. In conclusion, the preceding discussion highlights that the na\"ive factorization approach offers a satisfactory description for \DKpipi{} decays concerning the $P$-wave, albeit its prediction remains highly sensitive to the values of the Wilson coefficients. To determine the precise $a_{1,2}$ values, we propose a fitting procedure that will be feasible once we incorporate the $S$-wave contribution in the following subsection.

\subsection{Complete description}

To close our discussion and to extract our final results for the Wilson coefficients, we consider the full contribution, consisting on the $S$- and $P$-waves contributions. For later convenience, we consider independent Wilson coefficients $a_{1,2}^{S(P)}$ for the $S$- and $P$-waves, obtaining for the full BR in $10^{-2}$ units,
\begin{align}  
    \textrm{BR} =&{} 2.87 (a_1^P)^2 + 10.55 (a_2^P)^2 + 10.98 a_1^P a_2^P
   +0.92 (a_1^S)^2 + 1.46 (a_2^S)^2 -2.04 a_1^S a_2^S \nonumber\\ &{}
   -a_1^Pa_1^S( 0.01 \cos\delta -0.23\sin\delta) 
   +a_1^Pa_2^S( 0.06 \cos\delta -0.27\sin\delta)  \nonumber\\ &{}
   -a_2^Pa_1^S( 0.12 \cos\delta -0.49\sin\delta) 
   +a_2^Pa_2^S( 0.24 \cos\delta -0.58\sin\delta)\, , 
\end{align}
where a relative phase between the $S$- and $P$-waves has been introduced as well. 
For $a_i^S = a_i^P$ and $\delta=0$, that would correspond to our initial naive expectations, it is impossible to reproduce both the total BR (that amounts to $9.38(16)\%$ PDG~\cite{ParticleDataGroup:2022pth}) and the $P$-wave contribution simultaneously for $a_{1,2}$ values within the benchmark estimates.
Beyond this, the Dalitz-plot distribution shows an interference pattern amongst the $S$- and $P$-waves that necessarily requires an additional relative phase, as already observed in Refs.~\cite{Aitala:2002kr,Boito:2009qd}. This is a clear indication of non-factorizable effects~\cite{Cheng:2002wu,Qin:2013tje} for the $S$-wave contribution, the origin of which we speculate about later in our analysis. In the following, and abandoning the most conservative \nf{} approach, we conisder the possibility that such effects can still be effectively incorporated by adopting complex Wilson coefficients, which is nevertheless a common practice in the field, see Refs.~\cite{Cheng:2002wu,Qin:2013tje,Dedonder:2021dmb} and references therein. To do so, we take independent Wilson coefficients $a_{i}^{P}$ and $a_i^S e^{i\delta}$, with $a_i^{S,P}$ real, that can be physically interpreted as assigning different Wilson coefficients for the $D^+ \to \bar{K}^*\pi^+$ and $D^+ \to \bar{K}_0^*\pi^+$ subprocesses. Also, to help stabilizing the fit and try to keep the Wilson coefficients not far from benchmark estimates, it is useful to use priors in the fitting procedure.\footnote{In particular, our results assume Gaussian priors with $a_1^P=1.31(8)$, $a_1^S=1.31(31)$, $a_2^{S,P}=-0.55(30)$, $\delta=118(24)^{\circ}$, see also \cref{sec:WCs}.}
 Employing a Monte Carlo procedure that assumes Gaussian noise for the data and priors to fully account for correlations, we find\footnote{The uncertainties correspond to data, $S$- and $P$-wave model uncertainties and $F^{D\pi}$ form factors uncertainties. Whenever any source of uncertainty is irrelevant for a given parameter, this is omitted.}
\begin{align}
   a_1^P &{}= 1.40(8)_{\textrm{data}}(3)_S(1)_P[9], & 
   a_2^P &{}= -0.43(4)_{\textrm{data}}(2)_S(3)_P(2)_{F^{D\pi}}[6], \\ 
   a_1^S &{}= 2.04(17)_{\textrm{data}}(23)_S(1)_P[29], & 
   a_2^S &{}= -0.80(15)_{\textrm{data}}(1)_S(1)_P(5)_{F^{D\pi}}[16],
   \end{align}
\begin{equation}
   \delta = (116(3)_{\textrm{data}}(2)_S(1)_P[4])^{\circ}, 
\end{equation}
where the quantity in brackets refers to the total uncertainty, that is obtained by combining the individual ones in quadrature. The fit yields $\textrm{BR}=9.13(5)_{\textrm{data}}(1)_S(0)_P[5]\%$, a $P$-wave $\textrm{BR}=0.94(4)_{\textrm{data}}(3)_{S}[5]\%$, and the Dalitz-plot and invariant mass distribution shown in \cref{fig:diffDKpipi,fig:DalitzPlotDKpipi}. The description reproduces well all quantities and provides a reasonable approximation to first order, even if the fine details of the invariant mass distribution are not precisely reproduced (that is partly related to the $1.5\sigma$ deviation displayed by the total BR). This is nevertheless to be expected given that accurate descriptions of the Dalitz-plot seem to require, at this precision, non-resonant $I=2$ and $3/2$ contributions~\cite{Aitala:2002kr,CLEO:2008jus,Nakamura:2015qga,Niecknig:2015ija}, that are beyond our approach. While these can be incorporated in more sophisticated frameworks, this is commonly at the expense of abandoning the underlying microscopic theory. 
In the following, we discuss some aspects regarding the value found for the Wilson coefficients and the na{\"i}ve factorization hypothesis.
The result for $a_i^P$ notably lies within benchmark estimates, signifying an excellent performance of na{\"i}ve factorization (it is worth reiterating that such outcome critically relies on the predicted relative signs between the $a_{1,2}$ contributions). By contrast, this is not the case for the $a_i^S$ coefficients. Furthermore, the latter require an additional overall phase, that could be questioned on the basis of analyticity and unitarity, despite ad-hoc complex phases being common in phenomenological descriptions of $D$ decays~\cite{Cheng:2002wu,Qin:2013tje,Dedonder:2021dmb}. In our perspective, this additional phase can be attributed to genuine three-body effects, that are beyond the na{\"i}ve factorization hypothesis and can alter the original $S$-wave phase, see for instance \cite{Niecknig:2015ija}. The reason for such effects to be stronger for the $S$ wave might be its flat behavior, as compared to the $P$ wave that is fully-dominated by the $\bar{K}^*(892)$ contribution, that could make rescattering effects relevant along the entire spectra. It is important to note that while our phase is not dynamical, it can be viewed as its average value in the vicinity of the $\bar{K}^*(892)$ region, where the $S$ and $P$ waves exhibit significant interference. Indeed, the observed positive shift in this window is similar to the one found in Ref.~\cite{Niecknig:2015ija}.
Summarizing, it seems that strict \nf{} exhibits a successful performance in determining the $P$-wave contribution in this process, whereas the $S$ wave receives sizeable non-factorizable effects which can be nonetheless effectively accommodated in this picture by introducing complex Wilson coefficients, a common procedure found in the literature. We have argued that such coefficients can be attributed to genuine three-body effects and the absence of quasi-two body dynamics for the $S$ wave.
To further test this hypothesis and to better constraint the Wilson coefficients (that, for the $a_i^S$ cases largely depend on the assumed priors, see \cref{sec:WCs}), we propose studying \DsKKpi{} decays, that were not discussed in Ref.~\cite{Boito:2009qd} and can provide further insight in this respect. 

\begin{figure}
    \centering
    \includegraphics[width=0.85\textwidth]{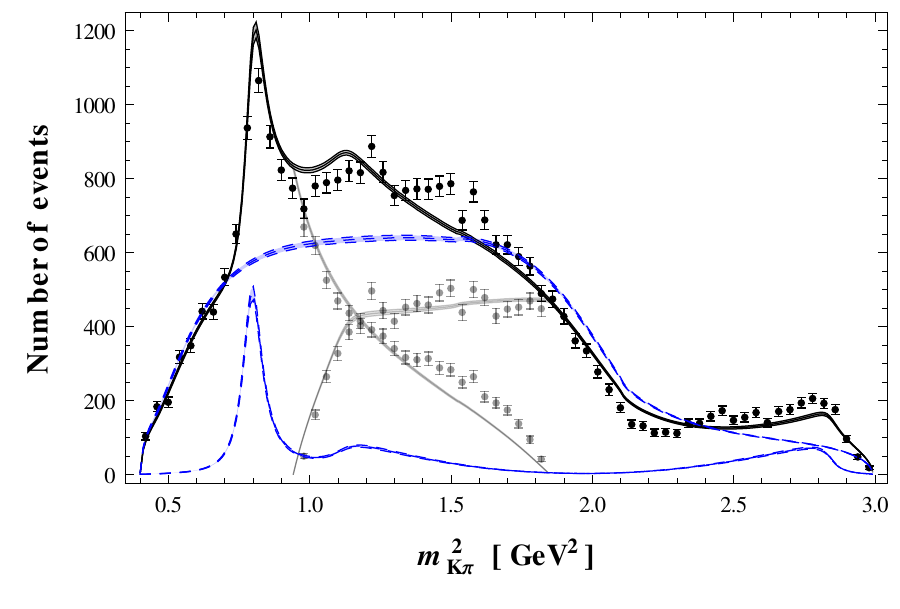}
    \caption{Differential decay width for \DKpipi{} compared to E791 data~\cite{Aitala:2002kr}. The dark gray band represents our model, while the light gray bands represent the low- and high-mass parts of the spectra. The dashed blue lines represent the scalar and vector components in our model. The bands are dominated by our MC fit uncertainties, but do not contain inherent uncertainties from the na{\"i}ve factorization hypothesis.}
    \label{fig:diffDKpipi}
\end{figure}
\begin{figure}
    \centering
    \includegraphics[width=0.495\textwidth]{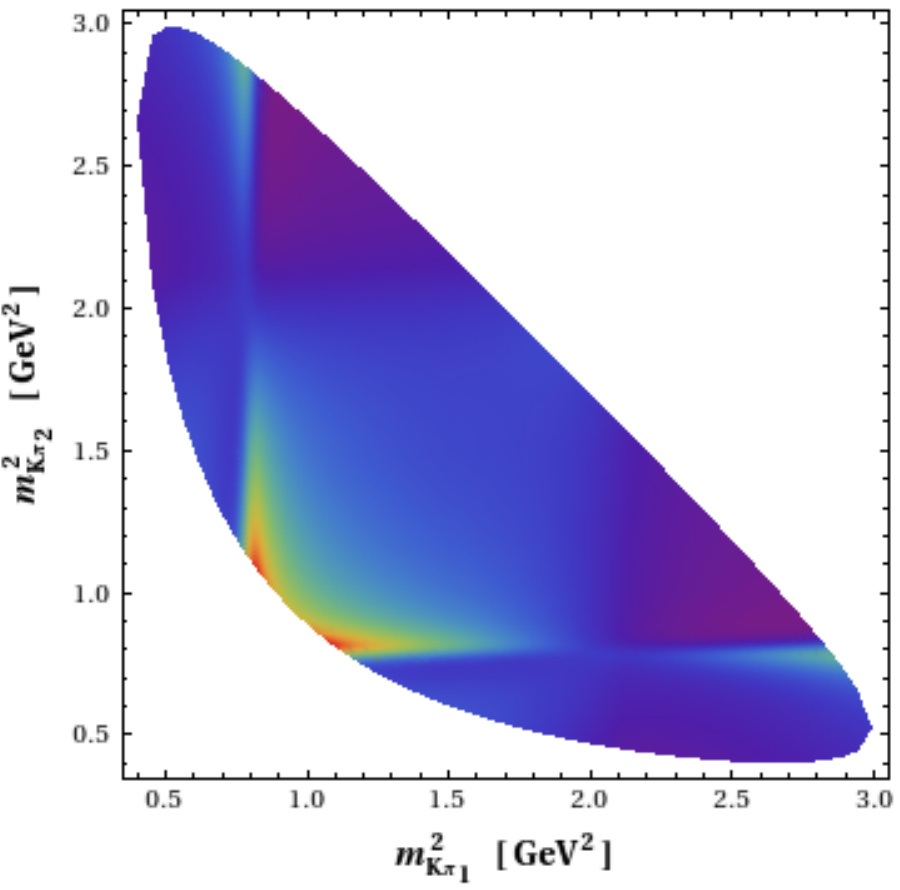}
    \includegraphics[width=0.495\textwidth]{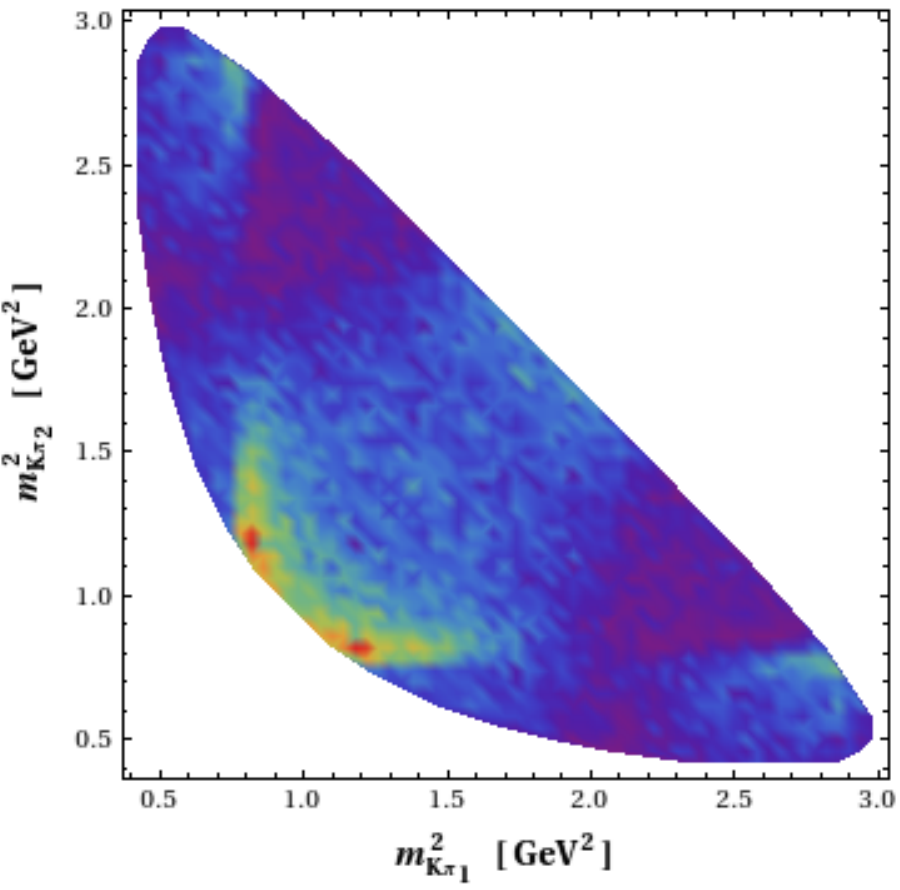}
    \caption{The symmetrized Dalitz plot for \DKpipi{} in our model (left) together with the experimental one from E791~\cite{Aitala:2002kr} (right).}
    \label{fig:DalitzPlotDKpipi}
\end{figure}

\section{Case II: \DsKKpi{} decays \label{sec:HDs}}

One advantage of the na{\"i}ve factorization approach over more sophisticated approaches lies in its connection with the underlying microscopic theory. This connection eanbles us to relate \DKpipi{} to \DsKKpi{} decays, partially accounting for $SU(3)$-breaking effects, that allows to test our concluding hypothesis in the previous section. In particular, the connection from  \DKpipi{} to \DsKKpi{} decays is achieved through the following replacements
\begin{align}
    f_{\pi} \to f_K\,, \quad 
    F_{+(0)}^{D\pi}(s) \to F_{+(0)}^{D_s K}(s)\,, \quad 
    F_4^{D_{\ell4}}(m_{\pi}^2,s,t) \to F_4^{D_{s\ell4}}(m_K^2,s,t)\,,
\end{align}
as well as $V_{ud}V_{cs}^* \to V_{us}V_{cd}^*$, $m_{D^+}\to m_{D_s}$ and $m_{K^{\pm}} \leftrightarrow m_{\pi^{\pm}}$ where necessary. We take $f_K/f_{\pi}=1.193(2)$~\cite{ParticleDataGroup:2022pth}, $F_{+(0)}^{D_s K}(0)=0.720(84)(13)$~\cite{Ablikim:2018upe} as well as effective masses $m_{D_{(0)}^{*0}}$. Regarding the semileptonic form factors, there are results in Refs.~\cite{Ablikim:2018upe} that show a similar pattern for the relative strengths, but do not report the overall normalization. We assume it to be the same based on approximate $U$-spin symmetry. With our model above,  and taking our results from previous section, we find  $\textrm{BR}=0.71(22)_{\textrm{MC}}(14)_{F^{D_sK}}[26]\times{10^{-4}}$, 
while for the $P$-wave one we find a fit fraction $0.29(10)_{\textrm{MC}}(2)_{F^{D_sK}}[11]$ (the invariant mass distribution is also shown as a 
light gray band in \cref{fig:DKKpiPlots}). Compared to the experimental result for the total $\textrm{BR} = 1.27(3)\times10^{-4}$~\cite{ParticleDataGroup:2022pth} and for the $P$-wave fit fraction, $0.47(22)(15)[27]$~\cite{BaBar:2010wqe}, we find seemingly lower values, albeit once more, within large uncertainties. Still, since we found that $a_i^S$ values had a sizeable dependence on the assumed priors, it is still possible to reproduce both decays at once. To explore this possibility and to better constraint the Wilson coefficients, we propose a combined fit that incorporates the total BR for \DsKKpi{}, obtaining 
\begin{align}
   a_1^P &{}= 1.37(7)_{\textrm{data}}(6)_S(1)_P(2)_{F^{D_sK}}[9], &
   a_2^P &{}= -0.42(4)_{\textrm{data}}(4)_S(3)_P(2)_{F^{D\pi}}(1)_{F^{D\pi}}[7], \\ 
   a_1^S &{}= 2.41(2)_{\textrm{data}}(26)_S(0)_P(7)_{F^{D_sK}}[27], &
   a_2^S &{}= -0.48(1)_{\textrm{data}}(3)_{F^{D\pi}}(6)_{F^{D_SK}}[7], 
\end{align}
\begin{equation}
   \delta = (119(3)_{\textrm{data}}(2)_S[4])^{\circ}.
\end{equation}
Regarding \DKpipi{} decays, these values imply $\textrm{BR}=9.14(5)_{\textrm{data}}(1)_P[5]\%$, and  $P$-wave $\textrm{BR}=0.93(4)_{\textrm{data}}(4)_{S}[6]\%$, essentially the same as in previous section}}. Regarding \DsKKpi{} decays, the new fit drastically improves the results. In particular, we obtain $\textrm{BR}=1.27(3)_{\textrm{data}}(1)_S(0)_P[3]\times10^{-4}$, in excellent agreement, while the $P$-wave fit fraction is reduced to $0.14(1)_{\textrm{data}}(3)_S(2)_P(2)_{F^{D\pi}}(4)_{F^{D_sK}}[6]$, in agreement with experiment at $1.2\sigma$ and suggesting a lower value. The Dalitz plot and invariant mass distribution for \DKpipi{} decays remain the same, while the corresponding quantities for \DsKKpi{} decays are shown in \cref{fig:DKKpiPlots}. It is remarkable that the interference pattern in the Dalitz 
plot is in excellent agreement with recent LHCb results~\cite{Aaij:2018hik}, that confirms once more an overall phase around $120^{\circ}$ among the $S$- and $P$-waves and reinforces our unified model. In addition, the invariant mas distribution seems in good agreement with the one in Ref.~\cite{BaBar:2010wqe}, that is subject to large uncertainties. Unfortunately, those datasets are not available. 
Summarizing, \DKpipi{} and \DsKKpi{} decays support a good performance of na{\"i}ve factorization regarding the $P$-wave contribution, that amounts to a quasi two body decay. This is not the case for the $S$-wave contribution, that seem to require sizeable non-factorizable effects. These can be nevertheless effectively encapsualted within the \nf{} approach through modified scalar Wilson coefficients, that we ascribed to genuine three-body effects and the absence of quasi-two body dynamics in the scalar channel.
In order to strengthen our findings, it would be interesting to have a detailed Dalitz plot analysis from Ref.~\cite{Aaij:2018hik}, that would allow to test our predictions for differential quantities. Likewise, we have large uncertainties from the current determination of $F_{+(0)}^{D_s K}(0)$. It would help reducing the current uncertainty if future lattice calculations become available. Similarly, it would be very interesting to have some higher-statistic measurement of semileptonic $D_s^+\to K^+\pi^-\ell^+\nu$ and $D^+\to K^-\pi^+\mu^+\nu$ decays, to better understand the involved form factors and to better judge the quality of our current description.
\\

\begin{figure}
    \centering
    \includegraphics[width=0.58\textwidth]{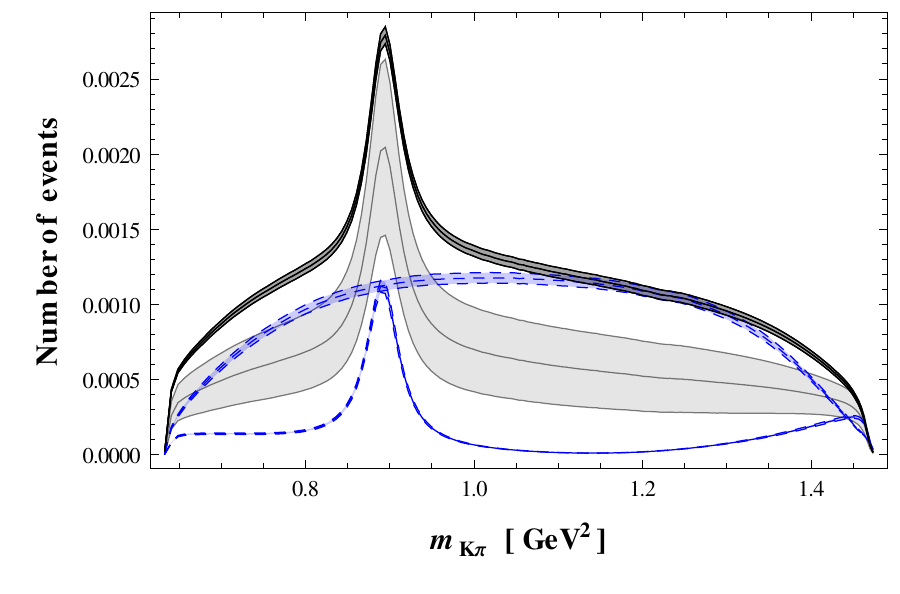}
    \includegraphics[width=0.41\textwidth]{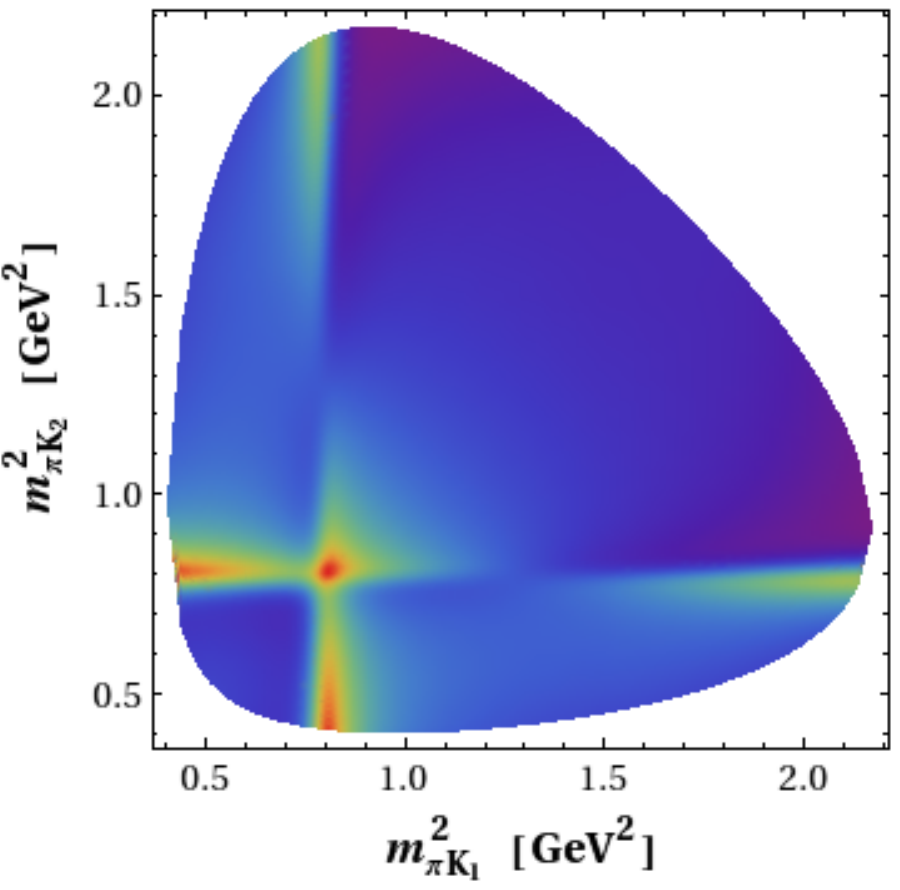}
    \caption{Left: invariant mass distribution for \DsKKpi{} decays. Our predictions from \DKpipi{} decays are shown as light-gray bands, whereas our fit result is shown as a gray band. The $S$-wave and $P$-wave contributions are shown as dahsed-blue bands. Right: Dalitz plot. Note in particular the depleted lower-left corner that requires the same relative scalar phase as in \DKpipi{} decays.}
    \label{fig:DKKpiPlots}
\end{figure}

\section{Conclusions\label{sec:conc}}

In this work we have performed a study of \DKpipi{} and \DsKKpi{} decays using the na{\"i}ve factorization framework, following the work in Ref.~\cite{Boito:2009qd}. Compared to that work, we have taken advantage of the precise data that has become available for semileptonic \DKpil{} decays~\cite{Ablikim:2015mjo}.
To that purpose, we have adopted a parametrization that incorporates final-state $K\pi$ interactions and fulfills analyticity and unitarity constraints below higher inelasticities, finding differences compared to Ref.~\cite{Boito:2009qd}, particularly for the $S$-wave. 
The resulting parametrization may be interesting in itself for experimentalists, and we encourage them to adopt it in future studies. 

Armed with these results, that were missing in Ref.~\cite{Boito:2009qd}, we have revisited the description of \DKpipi{} decays. Compared to Ref.~\cite{Boito:2009qd}, with our new paramtrization we are in the position to infer the relevant Wilson coefficients from \DKpipi{} decays. Remarkably, the $P$-wave contribution was accurately reproduced for benchmark values of the Wilson coefficients, a critical outcome given its strong dependence on the relative sign of $a_{1,2}$ contributions, which is predicted within our framework. Interestingly, the result is highly sensitive to the Wilson coefficients. 
By contrast, the $S$-wave contribution requires substantial deviations from benchmark values, as well as a complex phase. Whilst in our our opinion this represents a departure from the strict factorization framework, it effectively provides a reasonable description and is nevertehless common to phenomenological descriptions of $D$ decays. We attribute this to genuine three-body effects, beyond the capabilities of na{\"i}ve factorization, and to the absence of an effective quasi-two body description for the $S$-wave channel. 

To test our hypothesis and to better constrain the Wilson coefficients, we have investigated their counterpart in $D_s$ mesons decays, specifically, \DsKKpi{} decays, that were not explored in Ref~\cite{Boito:2009qd}. Notoriously, the results confirm the possibility to have a combined description for both decays. Our findings reaffirm that the $P$-wave contribution is successfully captured by the na{\"i}ve factorization framework, whereas the $S$-wave can again be effectively described by adopting complex Wilson coefficients. Remarkably, the overall phase, which is predicted from \DKpipi{} decays, successfully predicts the interference pattern observed in the Dalitz plot of \DsKKpi{} decays.

In the future, it would be interesting to have a available Dalitz-plot analysis, for which LHCb has accumulated data~\cite{Aaij:2018hik}. In addition, further results on semileptonic decays (possibly with muons), as well as $D_{(s)}\to \pi(K)$ form factors would allow to further test our results.

\section*{Acknowledgments}

P.~S.-P. thanks J.~Nieves for discussions on semileptonic form factors. This work has been supported by the European Union’s Horizon 2020 Research and Innovation Programme 
 under grants 754510
(EU, H2020-MSCA-COFUND2016) and 824093 (H2020-INFRAIA-2018-1), the Spanish Ministry of Science and Innovation under grants PID2020-112965GB-I00 and
PID2020-114767GB-I00, Junta de Andalucía under grants POSTDOC\_21\_00136 and
FQM-225, and the Secretaria d’Universitats i Recerca del Departament d’Empresa i Coneixement de la Generalitat de Catalunya under grant 2021 SGR 00649. IFAE is partially funded by the CERCA program of the Generalitat de Catalunya.

\appendix

\section{Definitions for \DKpil{} decays}\label{app:DSLdecays}

\subsection{Phase space and kinematics}

For this process, we take the conventions in Ref.~\cite{Kampf:2018wau}. Note in particular that our lepton-hadron-plane angle ($\phi$ in the following) defined in \cref{fig:Dl4kin} has opposite sign to that in Refs.~\cite{Lee:1992ih,Ablikim:2015mjo} ($\chi$ in the following).
  \begin{figure}[t]
      \centering
      \includegraphics[width=0.49\textwidth]{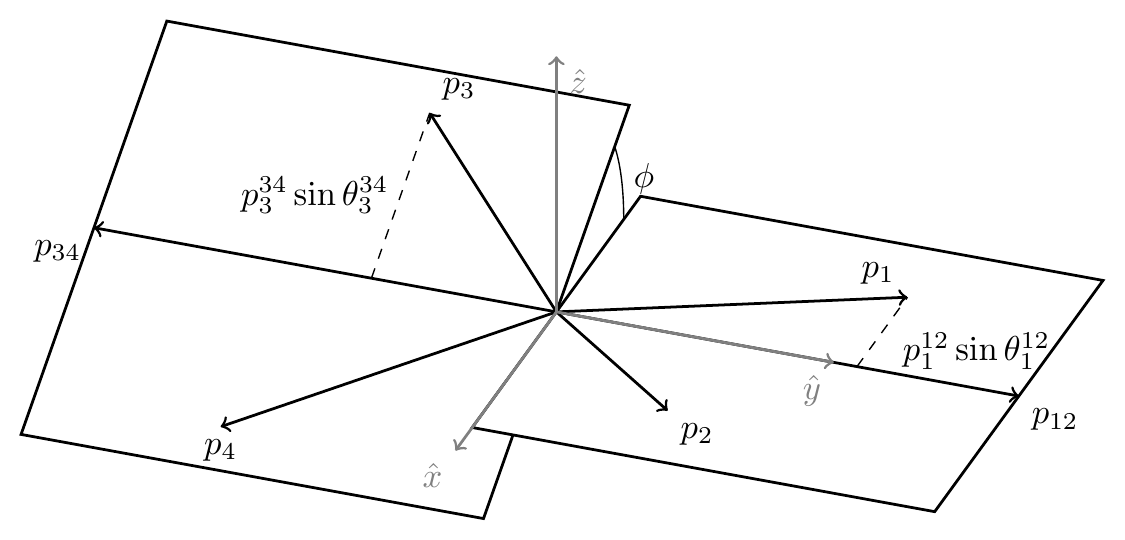}
      \includegraphics[width=0.49\textwidth]{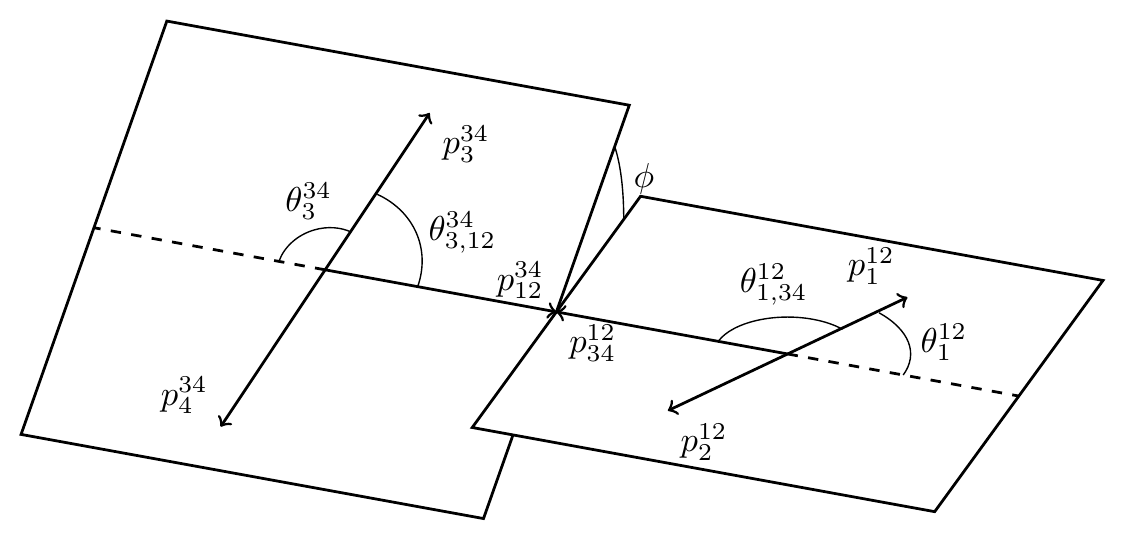}
      \caption{Definitions for the phase space variables in \DKpil{} decays. The particle labeling reads $\{1,2,3,4\} =\{K^-,\pi^+,\ell^+,\nu\}$.}
      \label{fig:Dl4kin}
  \end{figure}
The phase space can be described in terms of the invariant masses $p_{K\pi(\ell\nu)}^2=s_{K\pi(\ell\nu)}$, angles in the hadronic/leptonic reference frames $\theta_{K\pi(\ell\nu)}$ and hadron-lepton planes angle. For the calculation all that is required is ($p_{ij} = p_i +p_j$, $\bar{p}_{ij}=p_i -p_j$)
\begin{align}
    p_{K\pi}\cdot p_{\ell\nu} =&{} (m_D^2 -s_{K\pi} -s_{\ell\nu})/2 \equiv z \,,\\[1ex]
    \bar{p}_{K\pi}\cdot p_{\ell\nu} =&{} \tilde{\Delta}_{K\pi}z +X\beta_{K\pi}\cos\theta_{K\pi}\equiv \zeta \,,\\[1ex]
    p_{K\pi}\cdot \bar{p}_{\ell\nu} =&{} \tilde{\Delta}_{\ell\nu}z +X\beta_{\ell\nu}\cos\theta_{\ell\nu} \,,\\
    \bar{p}_{K\pi}\cdot \bar{p}_{\ell\nu} =&{} \Big[
      z\left( \tilde{\Delta}_{K\pi}\tilde{\Delta}_{\ell\nu} +\beta_{K\pi}\beta_{\ell\nu}\cos\theta_{K\pi}\cos\theta_{\ell\nu} \right) \nonumber\\ &{}
      +X\left( \tilde{\Delta}_{K\pi}\beta_{\ell\nu}\cos\theta_{\ell\nu}+\tilde{\Delta}_{\ell\nu}\beta_{K\pi}\cos\theta_{K\pi}\big)
    \right] \nonumber\\ 
    &{} -\sqrt{s_{K\pi}s_{\ell\nu}}\beta_{K\pi}\beta_{\ell\nu}\sin\theta_{K\pi}\sin\theta_{\ell\nu}\cos\phi \,,\\[1ex]
    \epsilon^{p_{K\pi}\bar{p}_{K\pi}p_{\ell\nu}\bar{p}_{\ell\nu}} =&{} -X\sqrt{s_{K\pi} s_{\ell\nu}}\beta_{K\pi}\beta_{\ell\nu}\sin\theta_{K\pi}\sin\theta_{\ell\nu}\sin\phi \,,
\end{align}
where $\tilde{\Delta}_{ij}=(p_i^2-p_j^2)/p_{ij}^2$, $\beta_{ij}=\lambda_{ij}^{1/2}/p_{ij}^2$, $X=\lambda_{K\pi,\ell\nu}^{1/2}/2$, and $\lambda_{ij}=[ p_{ij}^2 - (p_i^2 +p_j^2) ]^2 -4p_i^2p_j^2 $. Finally, the differential phase space can be defined as 
\begin{equation}
    d\Phi_4 = \frac{1}{(4\pi)^6}\frac{1}{2m_D^2} X\beta_{K\pi}\beta_{\ell\nu} ds_{K\pi}ds_{\ell\nu}d\cos\theta_{K\pi}d\cos\theta_{\ell\nu}d\phi \,.
\end{equation}

\subsection{Decay width}\label{sec:SLDW}

Following \cref{eq:Dl4M} and the notation for the hadronic form factors in \cref{eq:DKpi,eq:F1trans,eq:F2trans,eq:F3trans,eq:F4trans}, the differential decay width is given by
\begin{equation}
    d\Gamma = \frac{G_F^2 |V_{cs}|^2}{(4\pi)^6 m_D^3} X\beta_{K\pi}\beta_{\ell\nu}(H^{\mu\nu}L_{\mu\nu})
             ds_{K\pi}ds_{\ell\nu}d\cos\theta_{K\pi}d\cos\theta_{\ell\nu}d\phi \,.
\end{equation}
Taking in parallel to Ref.~\cite{Lee:1992ih} the following decomposition\footnote{Note in particular the minus sign in the $I_{7-9}$ terms due to our $\phi$ definition.} (for corresponding $CP$-related $D^-$ decays, $\phi\to-\phi$ needs to be taken)
\begin{multline}
    H^{\mu\nu}L_{\mu\nu} \equiv I_1 +I_2 \cos2\theta_{\ell\nu} +I_3 \sin^2\theta_{\ell\nu}\cos2\phi 
    +I_4 \sin2\theta_{\ell\nu}\cos\phi +I_5 \sin\theta_{\ell\nu}\cos\phi \\ +I_6\cos\theta_{\ell\nu}
    -I_7 \sin\theta_{\ell\nu}\sin\phi - I_8 \sin2\theta_{\ell\nu}\sin\phi- I_9 \sin^2\theta_{\ell\nu}\sin2\phi\,,
\end{multline}
the results in Ref.~\cite{Lee:1992ih} are modified for finite lepton masses ($m_{\nu}=0$) as follows 
\begin{align}
    I_1 ={}& \frac{1}{4}\beta_{\ell\nu}\left[ ( 1+\frac{m_{\ell}^2}{s_{\ell\nu}} )|F_1|^2 
          +\frac{3}{2}\sin^2\theta_{K\pi}( 1+\frac{m_{\ell}^2}{3s_{\ell\nu}} )(|F_2|^2 +|F_3|^2) + 
          \frac{2m_{\ell}^2}{s_{\ell\nu}}|F_4|^2\right],\label{eq:I1}\\
    I_2 ={}& -\frac{1}{4}\beta_{\ell\nu}^2\left[  |F_1|^2 
              -\frac{1}{2}\sin^2\theta_{K\pi}(|F_2|^2 +|F_3|^2) \right],\label{eq:I2}\\    
    I_3 ={}& -\frac{1}{4}\beta_{\ell\nu}^2\left[|F_2|^2 -|F_3|^2 \right]\sin^2\theta_{K\pi}\,,\label{eq:I3}\\
    I_4 ={}& \frac{1}{2}\beta_{\ell\nu}^2\operatorname{Re}(F_1F_2^*)\sin\theta_{K\pi}\,,\label{eq:I4}\\
    I_5 ={}& \beta_{\ell\nu}\operatorname{Re}\left[ F_1F_3^*
             +\frac{m_{\ell}^2}{s_{\ell\nu}}F_4F_2^*\right]\sin\theta_{K\pi}\,,\label{eq:I5}\\
    I_6 ={}& \beta_{\ell\nu}\operatorname{Re}\left[ F_2F_3^* \sin^2\theta_{K\pi}
             -\frac{m_{\ell}^2}{s_{\ell\nu}}F_1F_4^*\right], \label{eq:I6}\\
    I_7 ={}& \beta_{\ell\nu}\operatorname{Im}\left[ F_1F_2^*
             +\frac{m_{\ell}^2}{s_{\ell\nu}}F_4F_3^*\right]\sin\theta_{K\pi}\,,\label{eq:I7}\\
    I_8 ={}& \frac{1}{2}\beta_{\ell\nu}^2\operatorname{Im}(F_1F_3^*)\sin\theta_{K\pi}\,,\label{eq:I8}\\
    I_9 ={}& -\frac{1}{2}\beta_{\ell\nu}^2\operatorname{Im}(F_2F_3^*)\sin^2\theta_{K\pi}\,, \label{eq:I9}   
\end{align}
which in the $m_{\ell}\to0$ coincides with that in Ref.~\cite{Lee:1992ih}. Note that the hadronic matrix element can also be expressed in terms of the $F_i$ form factors as ($\xi = \Delta_{K\pi} X +z\beta_{K\pi}\cos\theta_{K\pi}$)
\begin{multline}
\bra{K^-\pi^+} \bar{s}\gamma^{\mu}(1-\gamma^5)c\ket{D^+}  = 
    \frac{iF_1}{X} \left(p_{K\pi}^{\mu} -p_{\ell\nu}^{\mu}\frac{z}{s_{\ell\nu}} \right) 
    +\frac{iF_4}{s_{\ell\nu}}p_{\ell\nu}^{\mu} \\
   +\frac{iF_2}{\beta_{K\pi}\sqrt{s_{K\pi}s_{\ell\nu}}}\left[
       \left( \bar{p}_{K\pi}^{\mu} -p_{\ell\nu}^{\mu}\frac{\zeta}{s_{\ell\nu}} \right)
      -\frac{\xi}{X} \left(p_{K\pi}^{\mu} -p_{\ell\nu}^{\mu}\frac{z}{s_{\ell\nu}} \right)
   \right]
     - \frac{F_3}{\beta_{K\pi}X\sqrt{s_{K\pi}s_{\ell\nu}}}\epsilon^{\mu p_{\ell\nu} p_{K\pi} \bar{p}_{K\pi}}\,.
\end{multline}

\section{Definitions in \DKpipi{} decays}\label{app:DKpipi}

Following \cref{eq:DKpipiFact} and the notation in \cref{sec:model} and Ref.~\cite{Boito:2009qd}, the matrix element of this process can be expressed as $\mathcal{M}= -i\frac{G_F}{\sqrt{2}} V_{ud}V_{cs}^*[ \mathcal{M}(s,t)+\mathcal{M}(t,s) ]$, where
\begin{multline}\label{eq:DalitzM}
    \mathcal{M}(s,t) = \frac{-a_1 f_{\pi}}{1- m_{\pi}^2/m_{D_s}^2} \left[\chi_S^{\textrm{eff}}(m_D^2 -s)F_{0}^{D_{\ell4}}(s)
                -N(s)F_{+}^{D_{\ell4}}(s) \frac{1}{2}(\chi_B^{\textrm{eff}} +\frac{m_D^2 -s}{2}\chi_C^{\textrm{eff}})  \right] \\ 
    +a_2\left[ 
     \frac{(m_D^2 -m_{\pi}^2)(m_K^2 -m_{\pi}^2)}{s}F_0^{K\pi}(s)F_0^{D\pi}(s) 
     + N(s)F_+^{K\pi}(s)F_+^{D\pi}(s)
    \right],
\end{multline}
with $F_{0,+}^{K\pi,D\pi}(s)$ standing for the relevant scalar(vector) form factors as defined in Ref.~\cite{Boito:2009qd}, and $N(s)= t-u  -(m_D^2 -m_{\pi}^2)(m_K^2 -m_{\pi}^2)/s$ defined below \cref{eq:F4V}. 
Consequently, the differential decay width can be expressed in terms of the Dalitz variables as 
\begin{equation}
    d\Gamma = \frac{1}{2}\frac{1}{(2\pi)^3}\frac{1}{32m_D^3}\frac{G_F^2 |V_{ud}V_{cs}^*|^2}{2} |\mathcal{M}(s,t) + \mathcal{M}(t,s)|^2\,.
\end{equation}
In our work, we adopt $|V_{ud}V_{cs}^*| = 0.971(17)$~\cite{ParticleDataGroup:2022pth}, $G_F=1.1663787\times10^{-5}\textrm{GeV}^{-2}$, $\Gamma_{D^+}=6.33\times10^{-13}$~GeV.

\section{The vector form factor description\label{sec:vectorFF}}

The phase for the vector form factor is that of the following~\cite{Escribano:2014joa}
\begin{equation}
    \tilde{f}_{+}^{K\pi} = 
    \frac{m_{K^*}^{2} - \left(\frac{192\pi}{\sigma_{K\pi}^3}\frac{\gamma_{K^*}}{m_{K^*}}\right)H_{K\pi}(0) +\gamma s}{m_{K^*}^{2} -s -\left(\frac{192\pi}{\sigma_{K\pi}^3(m_{K^*}^{2})}\frac{\gamma_{K^*}}{m_{K^*}}\right)H_{K\pi}(s)} -
     \frac{\gamma s}{m_{K^{*\prime}}^{2} -s -\left(\frac{192\pi}{\sigma_{K\pi}^3(m_{K^{*\prime}}^{2})}\frac{\gamma_{K^{*\prime}}}{m_{K^{*\prime}}}\right)H_{K\pi}(s)}\,,
\end{equation}
with $\sigma_{K\pi}^2(s) = \lambda(s,m_K^2,m_{\pi}^2)/s^2$, where we used the Kahlén function $\lambda(a,b,c)=a^2 +b^2 +c^2 -2ab -2ac -2bc$, and with 
\begin{multline}
    H_{K\pi}(s) = \frac{1}{(4\pi)^2}\frac{1}{12}\left[  s\sigma_{K\pi}^2(s)\bar{B}_0(s;m_{\pi}^2,m_K^2) -\frac{s}{2}\ln\frac{m_{\pi}^2 m_K^2}{\mu^4} \right. \\ \left. -\frac{(\Sigma_{K\pi}^2 -\Delta_{K\pi}^2 -\frac{s\Sigma_{K\pi}}{2})\ln\frac{m_K^2}{m_{\pi}^2}}{\Delta_{K\pi}} +\left(\frac{2}{3}s -2\Sigma_{K\pi} \right)
    \right],
\end{multline}
with $\Delta_{K\pi} = m_K^2 -m_{\pi}^2$ and $\Sigma_{K\pi} = m_K^2 +m_{\pi}^2$. The function $\bar{B}$ is defined in terms of the 1-loop two-point function $\bar{B}(s,m_K^2,m_{\pi}^2) = B(s,m_K^2,m_{\pi}^2) -B(0,m_K^2,m_{\pi}^2)$ and reads
\begin{equation}
    \bar{B}(s,m_K^2,m_{\pi}^2) = \frac{1}{2}\left[
2 +\left(\frac{\Delta_{K\pi}}{s} -\frac{\Sigma_{K\pi}}{\Delta_{K\pi}}\right)\ln\frac{m_{\pi}^2}{m_K^2} +2\sigma_{K\pi}(s)\ln\left(\frac{\Sigma_{K\pi} +s\sigma_{K\pi} -s}{2m_K m_{\pi}}\right)
    \right].
\end{equation}
In order to match their poles position we use the same parameters $m_{K^{*}} = 0.94338(69)~\textrm{GeV}$,  $\gamma_{K^{*}}=0.06666(8)~\textrm{GeV}$ and $m_{K^{*\prime}} = 1.379(36)~\textrm{GeV}$,  $\gamma_{K^{*\prime}}=0.196(66)~\textrm{GeV}$. Concerning $\gamma$, we choose $\gamma=0$ instead of $\gamma = -0.034$ since BES-III finds no evidence for a $K^*(1410)$. Still, the model allows for an easy extension to study possible effects of the $K^*(1410)$.

\section{Some results from the fit\label{sec:WCs}}

\begin{figure}
    \centering
    \includegraphics[width=0.42\textwidth]{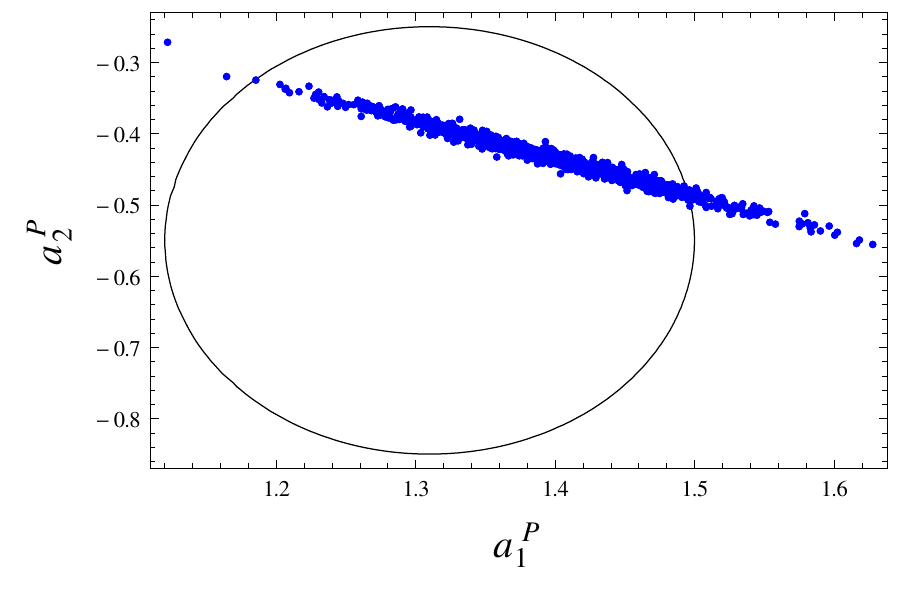}
    \includegraphics[width=0.42\textwidth]{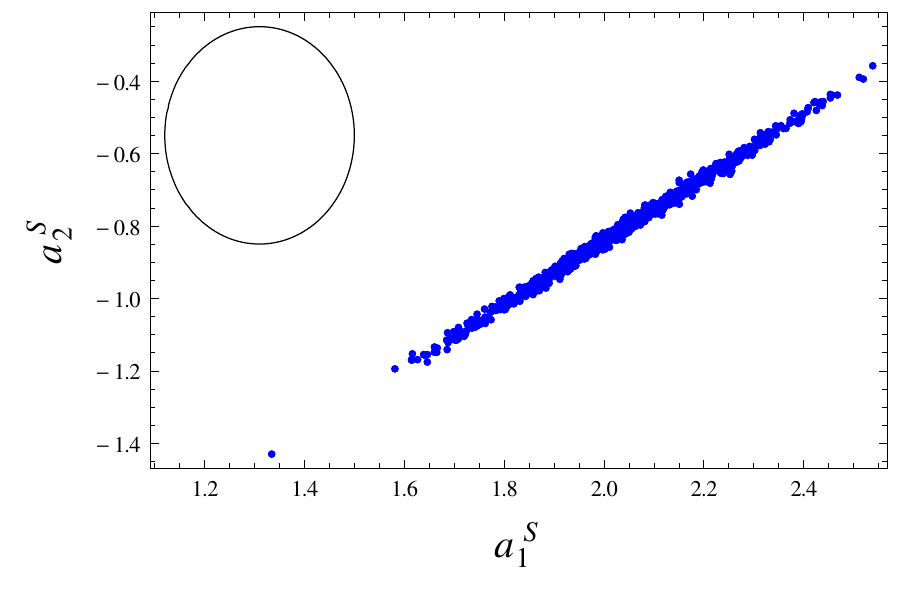}    
    \caption{The vector (left) and scalar (right) Wilson coefficients as obtained from our MC fit in \cref{sec:HD}. The blue points are the results from the MC fitting procedure with, whereas the ellipsis are the theoretically preferred values for the Wilson coefficients. Note in particular the strong correlations.}
    \label{fig:WCs}
\end{figure}

In \cref{fig:WCs}, we show our distribution of $a_1^{P,S}$~vs.~$a_2^{P,S}$ parameters obtained in our MC analysis. The benchmark values are shown as an ellipse. A strong correlation among them is clear. Note in particular that $a_i^P$ parameters fit well within benchmark estimates. This  allows to better constrain them using priors. On turn, $a_i^S$ values do not fit within the ellipse, that implies a large dependence on the assumed priors. This can be remedied by using $D_s$ decays.

\bibliographystyle{apsrev4-1}
\bibliography{references}

\begin{thebibliography}{42}%
\makeatletter
\providecommand \@ifxundefined [1]{%
 \@ifx{#1\undefined}
}%
\providecommand \@ifnum [1]{%
 \ifnum #1\expandafter \@firstoftwo
 \else \expandafter \@secondoftwo
 \fi
}%
\providecommand \@ifx [1]{%
 \ifx #1\expandafter \@firstoftwo
 \else \expandafter \@secondoftwo
 \fi
}%
\providecommand \natexlab [1]{#1}%
\providecommand \enquote  [1]{``#1''}%
\providecommand \bibnamefont  [1]{#1}%
\providecommand \bibfnamefont [1]{#1}%
\providecommand \citenamefont [1]{#1}%
\providecommand \href@noop [0]{\@secondoftwo}%
\providecommand \href [0]{\begingroup \@sanitize@url \@href}%
\providecommand \@href[1]{\@@startlink{#1}\@@href}%
\providecommand \@@href[1]{\endgroup#1\@@endlink}%
\providecommand \@sanitize@url [0]{\catcode `\\12\catcode `\$12\catcode
  `\&12\catcode `\#12\catcode `\^12\catcode `\_12\catcode `\%12\relax}%
\providecommand \@@startlink[1]{}%
\providecommand \@@endlink[0]{}%
\providecommand \url  [0]{\begingroup\@sanitize@url \@url }%
\providecommand \@url [1]{\endgroup\@href {#1}{\urlprefix }}%
\providecommand \urlprefix  [0]{URL }%
\providecommand \Eprint [0]{\href }%
\providecommand \doibase [0]{http://dx.doi.org/}%
\providecommand \selectlanguage [0]{\@gobble}%
\providecommand \bibinfo  [0]{\@secondoftwo}%
\providecommand \bibfield  [0]{\@secondoftwo}%
\providecommand \translation [1]{[#1]}%
\providecommand \BibitemOpen [0]{}%
\providecommand \bibitemStop [0]{}%
\providecommand \bibitemNoStop [0]{.\EOS\space}%
\providecommand \EOS [0]{\spacefactor3000\relax}%
\providecommand \BibitemShut  [1]{\csname bibitem#1\endcsname}%
\let\auto@bib@innerbib\@empty
\bibitem [{\citenamefont {Boito}\ and\ \citenamefont
  {Escribano}(2009)}]{Boito:2009qd}%
  \BibitemOpen
  \bibfield  {author} {\bibinfo {author} {\bibfnamefont {D.~R.}\ \bibnamefont
  {Boito}}\ and\ \bibinfo {author} {\bibfnamefont {R.}~\bibnamefont
  {Escribano}},\ }\href {\doibase 10.1103/PhysRevD.80.054007} {\bibfield
  {journal} {\bibinfo  {journal} {Phys. Rev.}\ }\textbf {\bibinfo {volume}
  {D80}},\ \bibinfo {pages} {054007} (\bibinfo {year} {2009})},\ \Eprint
  {http://arxiv.org/abs/0907.0189} {arXiv:0907.0189 [hep-ph]} \BibitemShut
  {NoStop}%
\bibitem [{\citenamefont {Ablikim}\ \emph {et~al.}(2016)\citenamefont {Ablikim}
  \emph {et~al.}}]{Ablikim:2015mjo}%
  \BibitemOpen
  \bibfield  {author} {\bibinfo {author} {\bibfnamefont {M.}~\bibnamefont
  {Ablikim}} \emph {et~al.} (\bibinfo {collaboration} {BESIII}),\ }\href
  {\doibase 10.1103/PhysRevD.94.032001} {\bibfield  {journal} {\bibinfo
  {journal} {Phys. Rev.}\ }\textbf {\bibinfo {volume} {D94}},\ \bibinfo {pages}
  {032001} (\bibinfo {year} {2016})},\ \Eprint
  {http://arxiv.org/abs/1512.08627} {arXiv:1512.08627 [hep-ex]} \BibitemShut
  {NoStop}%
\bibitem [{\citenamefont {Cheng}(2003)}]{Cheng:2002wu}%
  \BibitemOpen
  \bibfield  {author} {\bibinfo {author} {\bibfnamefont {H.-Y.}\ \bibnamefont
  {Cheng}},\ }\href {\doibase 10.1140/epjc/s2002-01065-6} {\bibfield  {journal}
  {\bibinfo  {journal} {Eur. Phys. J. C}\ }\textbf {\bibinfo {volume} {26}},\
  \bibinfo {pages} {551} (\bibinfo {year} {2003})},\ \Eprint
  {http://arxiv.org/abs/hep-ph/0202254} {arXiv:hep-ph/0202254} \BibitemShut
  {NoStop}%
\bibitem [{\citenamefont {Qin}\ \emph {et~al.}(2014)\citenamefont {Qin},
  \citenamefont {Li}, \citenamefont {L\"u},\ and\ \citenamefont
  {Yu}}]{Qin:2013tje}%
  \BibitemOpen
  \bibfield  {author} {\bibinfo {author} {\bibfnamefont {Q.}~\bibnamefont
  {Qin}}, \bibinfo {author} {\bibfnamefont {H.-n.}\ \bibnamefont {Li}},
  \bibinfo {author} {\bibfnamefont {C.-D.}\ \bibnamefont {L\"u}}, \ and\
  \bibinfo {author} {\bibfnamefont {F.-S.}\ \bibnamefont {Yu}},\ }\href
  {\doibase 10.1103/PhysRevD.89.054006} {\bibfield  {journal} {\bibinfo
  {journal} {Phys. Rev. D}\ }\textbf {\bibinfo {volume} {89}},\ \bibinfo
  {pages} {054006} (\bibinfo {year} {2014})},\ \Eprint
  {http://arxiv.org/abs/1305.7021} {arXiv:1305.7021 [hep-ph]} \BibitemShut
  {NoStop}%
\bibitem [{\citenamefont {Dedonder}\ \emph {et~al.}(2021)\citenamefont
  {Dedonder}, \citenamefont {Kami\'nski}, \citenamefont {Le\'sniak},\ and\
  \citenamefont {Loiseau}}]{Dedonder:2021dmb}%
  \BibitemOpen
  \bibfield  {author} {\bibinfo {author} {\bibfnamefont {J.~P.}\ \bibnamefont
  {Dedonder}}, \bibinfo {author} {\bibfnamefont {R.}~\bibnamefont
  {Kami\'nski}}, \bibinfo {author} {\bibfnamefont {L.}~\bibnamefont
  {Le\'sniak}}, \ and\ \bibinfo {author} {\bibfnamefont {B.}~\bibnamefont
  {Loiseau}},\ }\href {\doibase 10.1103/PhysRevD.103.114028} {\bibfield
  {journal} {\bibinfo  {journal} {Phys. Rev. D}\ }\textbf {\bibinfo {volume}
  {103}},\ \bibinfo {pages} {114028} (\bibinfo {year} {2021})},\ \Eprint
  {http://arxiv.org/abs/2105.03355} {arXiv:2105.03355 [hep-ph]} \BibitemShut
  {NoStop}%
\bibitem [{\citenamefont {Magalhaes}\ \emph {et~al.}(2011)\citenamefont
  {Magalhaes}, \citenamefont {Robilotta}, \citenamefont {Guimaraes},
  \citenamefont {Frederico}, \citenamefont {de~Paula}, \citenamefont {Bediaga},
  \citenamefont {Reis}, \citenamefont {Maekawa},\ and\ \citenamefont
  {Zarnauskas}}]{Magalhaes:2011sh}%
  \BibitemOpen
  \bibfield  {author} {\bibinfo {author} {\bibfnamefont {P.~C.}\ \bibnamefont
  {Magalhaes}}, \bibinfo {author} {\bibfnamefont {M.~R.}\ \bibnamefont
  {Robilotta}}, \bibinfo {author} {\bibfnamefont {K.~S. F.~F.}\ \bibnamefont
  {Guimaraes}}, \bibinfo {author} {\bibfnamefont {T.}~\bibnamefont
  {Frederico}}, \bibinfo {author} {\bibfnamefont {W.}~\bibnamefont {de~Paula}},
  \bibinfo {author} {\bibfnamefont {I.}~\bibnamefont {Bediaga}}, \bibinfo
  {author} {\bibfnamefont {A.~C.~d.}\ \bibnamefont {Reis}}, \bibinfo {author}
  {\bibfnamefont {C.~M.}\ \bibnamefont {Maekawa}}, \ and\ \bibinfo {author}
  {\bibfnamefont {G.~R.~S.}\ \bibnamefont {Zarnauskas}},\ }\href {\doibase
  10.1103/PhysRevD.84.094001} {\bibfield  {journal} {\bibinfo  {journal} {Phys.
  Rev. D}\ }\textbf {\bibinfo {volume} {84}},\ \bibinfo {pages} {094001}
  (\bibinfo {year} {2011})},\ \Eprint {http://arxiv.org/abs/1105.5120}
  {arXiv:1105.5120 [hep-ph]} \BibitemShut {NoStop}%
\bibitem [{\citenamefont {Guimar\~aes}\ \emph {et~al.}(2014)\citenamefont
  {Guimar\~aes}, \citenamefont {Louren\c{c}o}, \citenamefont {de~Paula},
  \citenamefont {Frederico},\ and\ \citenamefont {dos
  Reis}}]{Guimaraes:2014kor}%
  \BibitemOpen
  \bibfield  {author} {\bibinfo {author} {\bibfnamefont {K.~S. F.~F.}\
  \bibnamefont {Guimar\~aes}}, \bibinfo {author} {\bibfnamefont
  {O.}~\bibnamefont {Louren\c{c}o}}, \bibinfo {author} {\bibfnamefont
  {W.}~\bibnamefont {de~Paula}}, \bibinfo {author} {\bibfnamefont
  {T.}~\bibnamefont {Frederico}}, \ and\ \bibinfo {author} {\bibfnamefont
  {A.~C.}\ \bibnamefont {dos Reis}},\ }\href {\doibase 10.1007/JHEP08(2014)135}
  {\bibfield  {journal} {\bibinfo  {journal} {JHEP}\ }\textbf {\bibinfo
  {volume} {08}},\ \bibinfo {pages} {135} (\bibinfo {year} {2014})},\ \Eprint
  {http://arxiv.org/abs/1404.3797} {arXiv:1404.3797 [hep-ph]} \BibitemShut
  {NoStop}%
\bibitem [{\citenamefont {Magalh\~aes}\ and\ \citenamefont
  {Robilotta}(2015)}]{Magalhaes:2015fva}%
  \BibitemOpen
  \bibfield  {author} {\bibinfo {author} {\bibfnamefont {P.~C.}\ \bibnamefont
  {Magalh\~aes}}\ and\ \bibinfo {author} {\bibfnamefont {M.~R.}\ \bibnamefont
  {Robilotta}},\ }\href {\doibase 10.1103/PhysRevD.92.094005} {\bibfield
  {journal} {\bibinfo  {journal} {Phys. Rev. D}\ }\textbf {\bibinfo {volume}
  {92}},\ \bibinfo {pages} {094005} (\bibinfo {year} {2015})},\ \Eprint
  {http://arxiv.org/abs/1504.06346} {arXiv:1504.06346 [hep-ph]} \BibitemShut
  {NoStop}%
\bibitem [{\citenamefont {Nakamura}(2016)}]{Nakamura:2015qga}%
  \BibitemOpen
  \bibfield  {author} {\bibinfo {author} {\bibfnamefont {S.~X.}\ \bibnamefont
  {Nakamura}},\ }\href {\doibase 10.1103/PhysRevD.93.014005} {\bibfield
  {journal} {\bibinfo  {journal} {Phys. Rev. D}\ }\textbf {\bibinfo {volume}
  {93}},\ \bibinfo {pages} {014005} (\bibinfo {year} {2016})},\ \Eprint
  {http://arxiv.org/abs/1504.02557} {arXiv:1504.02557 [hep-ph]} \BibitemShut
  {NoStop}%
\bibitem [{\citenamefont {Niecknig}\ and\ \citenamefont
  {Kubis}(2015)}]{Niecknig:2015ija}%
  \BibitemOpen
  \bibfield  {author} {\bibinfo {author} {\bibfnamefont {F.}~\bibnamefont
  {Niecknig}}\ and\ \bibinfo {author} {\bibfnamefont {B.}~\bibnamefont
  {Kubis}},\ }\href {\doibase 10.1007/JHEP10(2015)142} {\bibfield  {journal}
  {\bibinfo  {journal} {JHEP}\ }\textbf {\bibinfo {volume} {10}},\ \bibinfo
  {pages} {142} (\bibinfo {year} {2015})},\ \Eprint
  {http://arxiv.org/abs/1509.03188} {arXiv:1509.03188 [hep-ph]} \BibitemShut
  {NoStop}%
\bibitem [{\citenamefont {Diakonou}\ and\ \citenamefont
  {Diakonos}(1989)}]{Diakonou:1989sf}%
  \BibitemOpen
  \bibfield  {author} {\bibinfo {author} {\bibfnamefont {M.}~\bibnamefont
  {Diakonou}}\ and\ \bibinfo {author} {\bibfnamefont {F.}~\bibnamefont
  {Diakonos}},\ }\href {\doibase 10.1016/0370-2693(89)91146-5} {\bibfield
  {journal} {\bibinfo  {journal} {Phys. Lett. B}\ }\textbf {\bibinfo {volume}
  {216}},\ \bibinfo {pages} {436} (\bibinfo {year} {1989})}\BibitemShut
  {NoStop}%
\bibitem [{\citenamefont {Oller}(2005)}]{Oller:2004xm}%
  \BibitemOpen
  \bibfield  {author} {\bibinfo {author} {\bibfnamefont {J.~A.}\ \bibnamefont
  {Oller}},\ }\href {\doibase 10.1103/PhysRevD.71.054030} {\bibfield  {journal}
  {\bibinfo  {journal} {Phys. Rev. D}\ }\textbf {\bibinfo {volume} {71}},\
  \bibinfo {pages} {054030} (\bibinfo {year} {2005})},\ \Eprint
  {http://arxiv.org/abs/hep-ph/0411105} {arXiv:hep-ph/0411105} \BibitemShut
  {NoStop}%
\bibitem [{\citenamefont {Buras}(1995)}]{Buras:1994ij}%
  \BibitemOpen
  \bibfield  {author} {\bibinfo {author} {\bibfnamefont {A.~J.}\ \bibnamefont
  {Buras}},\ }\href {\doibase 10.1016/0550-3213(94)00482-T} {\bibfield
  {journal} {\bibinfo  {journal} {Nucl. Phys.}\ }\textbf {\bibinfo {volume}
  {B434}},\ \bibinfo {pages} {606} (\bibinfo {year} {1995})},\ \Eprint
  {http://arxiv.org/abs/hep-ph/9409309} {arXiv:hep-ph/9409309 [hep-ph]}
  \BibitemShut {NoStop}%
\bibitem [{\citenamefont {Neubert}\ \emph {et~al.}(1992)\citenamefont
  {Neubert}, \citenamefont {Rieckert}, \citenamefont {Stech},\ and\
  \citenamefont {Xu}}]{Neubert:1991we}%
  \BibitemOpen
  \bibfield  {author} {\bibinfo {author} {\bibfnamefont {M.}~\bibnamefont
  {Neubert}}, \bibinfo {author} {\bibfnamefont {V.}~\bibnamefont {Rieckert}},
  \bibinfo {author} {\bibfnamefont {B.}~\bibnamefont {Stech}}, \ and\ \bibinfo
  {author} {\bibfnamefont {Q.~P.}\ \bibnamefont {Xu}},\ }\href {\doibase
  10.1142/9789814503587_0005} {\bibfield  {journal} {\bibinfo  {journal} {Adv.
  Ser. Direct. High Energy Phys.}\ }\textbf {\bibinfo {volume} {10}},\ \bibinfo
  {pages} {286} (\bibinfo {year} {1992})}\BibitemShut {NoStop}%
\bibitem [{\citenamefont {Workman}\ \emph {et~al.}()\citenamefont {Workman}
  \emph {et~al.}}]{ParticleDataGroup:2022pth}%
  \BibitemOpen
  \bibfield  {author} {\bibinfo {author} {\bibfnamefont {R.~L.}\ \bibnamefont
  {Workman}} \emph {et~al.} (\bibinfo {collaboration} {Particle Data Group}),\
  }\href {\doibase 10.1093/ptep/ptac097} {\bibfield  {journal} {\bibinfo
  {journal} {PTEP}\ }\textbf {\bibinfo {volume} {2022}},\ \bibinfo {pages}
  {083C01}}\BibitemShut {NoStop}%
\bibitem [{\citenamefont {Jamin}\ \emph {et~al.}(2006)\citenamefont {Jamin},
  \citenamefont {Oller},\ and\ \citenamefont {Pich}}]{Jamin:2006tj}%
  \BibitemOpen
  \bibfield  {author} {\bibinfo {author} {\bibfnamefont {M.}~\bibnamefont
  {Jamin}}, \bibinfo {author} {\bibfnamefont {J.~A.}\ \bibnamefont {Oller}}, \
  and\ \bibinfo {author} {\bibfnamefont {A.}~\bibnamefont {Pich}},\ }\href
  {\doibase 10.1103/PhysRevD.74.074009} {\bibfield  {journal} {\bibinfo
  {journal} {Phys. Rev.}\ }\textbf {\bibinfo {volume} {D74}},\ \bibinfo {pages}
  {074009} (\bibinfo {year} {2006})},\ \Eprint
  {http://arxiv.org/abs/hep-ph/0605095} {arXiv:hep-ph/0605095 [hep-ph]}
  \BibitemShut {NoStop}%
\bibitem [{\citenamefont {Boito}\ \emph {et~al.}(2010)\citenamefont {Boito},
  \citenamefont {Escribano},\ and\ \citenamefont {Jamin}}]{Boito:2010me}%
  \BibitemOpen
  \bibfield  {author} {\bibinfo {author} {\bibfnamefont {D.~R.}\ \bibnamefont
  {Boito}}, \bibinfo {author} {\bibfnamefont {R.}~\bibnamefont {Escribano}}, \
  and\ \bibinfo {author} {\bibfnamefont {M.}~\bibnamefont {Jamin}},\ }\href
  {\doibase 10.1007/JHEP09(2010)031} {\bibfield  {journal} {\bibinfo  {journal}
  {JHEP}\ }\textbf {\bibinfo {volume} {09}},\ \bibinfo {pages} {031} (\bibinfo
  {year} {2010})},\ \Eprint {http://arxiv.org/abs/1007.1858} {arXiv:1007.1858
  [hep-ph]} \BibitemShut {NoStop}%
\bibitem [{\citenamefont {Link}\ \emph {et~al.}(2002)\citenamefont {Link} \emph
  {et~al.}}]{Link:2002wg}%
  \BibitemOpen
  \bibfield  {author} {\bibinfo {author} {\bibfnamefont {J.~M.}\ \bibnamefont
  {Link}} \emph {et~al.} (\bibinfo {collaboration} {FOCUS}),\ }\href {\doibase
  10.1016/S0370-2693(02)02386-9} {\bibfield  {journal} {\bibinfo  {journal}
  {Phys. Lett.}\ }\textbf {\bibinfo {volume} {B544}},\ \bibinfo {pages} {89}
  (\bibinfo {year} {2002})},\ \Eprint {http://arxiv.org/abs/hep-ex/0207049}
  {arXiv:hep-ex/0207049 [hep-ex]} \BibitemShut {NoStop}%
\bibitem [{\citenamefont {Briere}\ \emph {et~al.}(2010)\citenamefont {Briere}
  \emph {et~al.}}]{Briere:2010zc}%
  \BibitemOpen
  \bibfield  {author} {\bibinfo {author} {\bibfnamefont {R.~A.}\ \bibnamefont
  {Briere}} \emph {et~al.} (\bibinfo {collaboration} {CLEO}),\ }\href {\doibase
  10.1103/PhysRevD.81.112001} {\bibfield  {journal} {\bibinfo  {journal} {Phys.
  Rev.}\ }\textbf {\bibinfo {volume} {D81}},\ \bibinfo {pages} {112001}
  (\bibinfo {year} {2010})},\ \Eprint {http://arxiv.org/abs/1004.1954}
  {arXiv:1004.1954 [hep-ex]} \BibitemShut {NoStop}%
\bibitem [{\citenamefont {Lee}\ \emph {et~al.}(1992)\citenamefont {Lee},
  \citenamefont {Lu},\ and\ \citenamefont {Wise}}]{Lee:1992ih}%
  \BibitemOpen
  \bibfield  {author} {\bibinfo {author} {\bibfnamefont {C.~L.~Y.}\
  \bibnamefont {Lee}}, \bibinfo {author} {\bibfnamefont {M.}~\bibnamefont
  {Lu}}, \ and\ \bibinfo {author} {\bibfnamefont {M.~B.}\ \bibnamefont
  {Wise}},\ }\href {\doibase 10.1103/PhysRevD.46.5040} {\bibfield  {journal}
  {\bibinfo  {journal} {Phys. Rev.}\ }\textbf {\bibinfo {volume} {D46}},\
  \bibinfo {pages} {5040} (\bibinfo {year} {1992})}\BibitemShut {NoStop}%
\bibitem [{\citenamefont {Bajc}\ \emph {et~al.}(1998)\citenamefont {Bajc},
  \citenamefont {Fajfer}, \citenamefont {Oakes},\ and\ \citenamefont
  {Pham}}]{Bajc:1997nx}%
  \BibitemOpen
  \bibfield  {author} {\bibinfo {author} {\bibfnamefont {B.}~\bibnamefont
  {Bajc}}, \bibinfo {author} {\bibfnamefont {S.}~\bibnamefont {Fajfer}},
  \bibinfo {author} {\bibfnamefont {R.~J.}\ \bibnamefont {Oakes}}, \ and\
  \bibinfo {author} {\bibfnamefont {T.~N.}\ \bibnamefont {Pham}},\ }\href
  {\doibase 10.1103/PhysRevD.58.054009} {\bibfield  {journal} {\bibinfo
  {journal} {Phys. Rev.}\ }\textbf {\bibinfo {volume} {D58}},\ \bibinfo {pages}
  {054009} (\bibinfo {year} {1998})},\ \Eprint
  {http://arxiv.org/abs/hep-ph/9710422} {arXiv:hep-ph/9710422 [hep-ph]}
  \BibitemShut {NoStop}%
\bibitem [{\citenamefont {Bernard}\ \emph {et~al.}(2006)\citenamefont
  {Bernard}, \citenamefont {Oertel}, \citenamefont {Passemar},\ and\
  \citenamefont {Stern}}]{Bernard:2006gy}%
  \BibitemOpen
  \bibfield  {author} {\bibinfo {author} {\bibfnamefont {V.}~\bibnamefont
  {Bernard}}, \bibinfo {author} {\bibfnamefont {M.}~\bibnamefont {Oertel}},
  \bibinfo {author} {\bibfnamefont {E.}~\bibnamefont {Passemar}}, \ and\
  \bibinfo {author} {\bibfnamefont {J.}~\bibnamefont {Stern}},\ }\href
  {\doibase 10.1016/j.physletb.2006.05.079} {\bibfield  {journal} {\bibinfo
  {journal} {Phys. Lett. B}\ }\textbf {\bibinfo {volume} {638}},\ \bibinfo
  {pages} {480} (\bibinfo {year} {2006})},\ \Eprint
  {http://arxiv.org/abs/hep-ph/0603202} {arXiv:hep-ph/0603202} \BibitemShut
  {NoStop}%
\bibitem [{\citenamefont {Bernard}\ \emph {et~al.}(2009)\citenamefont
  {Bernard}, \citenamefont {Oertel}, \citenamefont {Passemar},\ and\
  \citenamefont {Stern}}]{Bernard:2009zm}%
  \BibitemOpen
  \bibfield  {author} {\bibinfo {author} {\bibfnamefont {V.}~\bibnamefont
  {Bernard}}, \bibinfo {author} {\bibfnamefont {M.}~\bibnamefont {Oertel}},
  \bibinfo {author} {\bibfnamefont {E.}~\bibnamefont {Passemar}}, \ and\
  \bibinfo {author} {\bibfnamefont {J.}~\bibnamefont {Stern}},\ }\href
  {\doibase 10.1103/PhysRevD.80.034034} {\bibfield  {journal} {\bibinfo
  {journal} {Phys. Rev. D}\ }\textbf {\bibinfo {volume} {80}},\ \bibinfo
  {pages} {034034} (\bibinfo {year} {2009})},\ \Eprint
  {http://arxiv.org/abs/0903.1654} {arXiv:0903.1654 [hep-ph]} \BibitemShut
  {NoStop}%
\bibitem [{\citenamefont {Bernard}(2014)}]{Bernard:2013jxa}%
  \BibitemOpen
  \bibfield  {author} {\bibinfo {author} {\bibfnamefont {V.}~\bibnamefont
  {Bernard}},\ }\href {\doibase 10.1007/JHEP06(2014)082} {\bibfield  {journal}
  {\bibinfo  {journal} {JHEP}\ }\textbf {\bibinfo {volume} {06}},\ \bibinfo
  {pages} {082} (\bibinfo {year} {2014})},\ \Eprint
  {http://arxiv.org/abs/1311.2569} {arXiv:1311.2569 [hep-ph]} \BibitemShut
  {NoStop}%
\bibitem [{\citenamefont {Wirbel}\ \emph {et~al.}(1985)\citenamefont {Wirbel},
  \citenamefont {Stech},\ and\ \citenamefont {Bauer}}]{Wirbel:1985ji}%
  \BibitemOpen
  \bibfield  {author} {\bibinfo {author} {\bibfnamefont {M.}~\bibnamefont
  {Wirbel}}, \bibinfo {author} {\bibfnamefont {B.}~\bibnamefont {Stech}}, \
  and\ \bibinfo {author} {\bibfnamefont {M.}~\bibnamefont {Bauer}},\ }\href
  {\doibase 10.1007/BF01560299} {\bibfield  {journal} {\bibinfo  {journal} {Z.
  Phys.}\ }\textbf {\bibinfo {volume} {C29}},\ \bibinfo {pages} {637} (\bibinfo
  {year} {1985})}\BibitemShut {NoStop}%
\bibitem [{\citenamefont {Bernard}\ \emph {et~al.}(1992)\citenamefont
  {Bernard}, \citenamefont {El-Khadra},\ and\ \citenamefont
  {Soni}}]{Bernard:1991bz}%
  \BibitemOpen
  \bibfield  {author} {\bibinfo {author} {\bibfnamefont {C.~W.}\ \bibnamefont
  {Bernard}}, \bibinfo {author} {\bibfnamefont {A.~X.}\ \bibnamefont
  {El-Khadra}}, \ and\ \bibinfo {author} {\bibfnamefont {A.}~\bibnamefont
  {Soni}},\ }\href {\doibase 10.1103/PhysRevD.45.869} {\bibfield  {journal}
  {\bibinfo  {journal} {Phys. Rev. D}\ }\textbf {\bibinfo {volume} {45}},\
  \bibinfo {pages} {869} (\bibinfo {year} {1992})}\BibitemShut {NoStop}%
\bibitem [{\citenamefont {Bowler}\ \emph {et~al.}(1995)\citenamefont {Bowler},
  \citenamefont {Hazel}, \citenamefont {Hoeber}, \citenamefont {Kenway},
  \citenamefont {Richards}, \citenamefont {Lellouch}, \citenamefont {Nieves},
  \citenamefont {Sachrajda},\ and\ \citenamefont {Wittig}}]{Bowler:1994zr}%
  \BibitemOpen
  \bibfield  {author} {\bibinfo {author} {\bibfnamefont {K.~C.}\ \bibnamefont
  {Bowler}}, \bibinfo {author} {\bibfnamefont {N.~M.}\ \bibnamefont {Hazel}},
  \bibinfo {author} {\bibfnamefont {H.}~\bibnamefont {Hoeber}}, \bibinfo
  {author} {\bibfnamefont {R.~D.}\ \bibnamefont {Kenway}}, \bibinfo {author}
  {\bibfnamefont {D.~G.}\ \bibnamefont {Richards}}, \bibinfo {author}
  {\bibfnamefont {L.}~\bibnamefont {Lellouch}}, \bibinfo {author}
  {\bibfnamefont {J.}~\bibnamefont {Nieves}}, \bibinfo {author} {\bibfnamefont
  {C.~T.}\ \bibnamefont {Sachrajda}}, \ and\ \bibinfo {author} {\bibfnamefont
  {H.}~\bibnamefont {Wittig}} (\bibinfo {collaboration} {UKQCD}),\ }\href
  {\doibase 10.1103/PhysRevD.51.4905} {\bibfield  {journal} {\bibinfo
  {journal} {Phys. Rev. D}\ }\textbf {\bibinfo {volume} {51}},\ \bibinfo
  {pages} {4905} (\bibinfo {year} {1995})},\ \Eprint
  {http://arxiv.org/abs/hep-lat/9410012} {arXiv:hep-lat/9410012} \BibitemShut
  {NoStop}%
\bibitem [{\citenamefont {Richman}\ and\ \citenamefont
  {Burchat}(1995)}]{Richman:1995wm}%
  \BibitemOpen
  \bibfield  {author} {\bibinfo {author} {\bibfnamefont {J.~D.}\ \bibnamefont
  {Richman}}\ and\ \bibinfo {author} {\bibfnamefont {P.~R.}\ \bibnamefont
  {Burchat}},\ }\href {\doibase 10.1103/RevModPhys.67.893} {\bibfield
  {journal} {\bibinfo  {journal} {Rev. Mod. Phys.}\ }\textbf {\bibinfo {volume}
  {67}},\ \bibinfo {pages} {893} (\bibinfo {year} {1995})},\ \Eprint
  {http://arxiv.org/abs/hep-ph/9508250} {arXiv:hep-ph/9508250 [hep-ph]}
  \BibitemShut {NoStop}%
\bibitem [{\citenamefont {Melikhov}\ and\ \citenamefont
  {Stech}(2000)}]{Melikhov:2000yu}%
  \BibitemOpen
  \bibfield  {author} {\bibinfo {author} {\bibfnamefont {D.}~\bibnamefont
  {Melikhov}}\ and\ \bibinfo {author} {\bibfnamefont {B.}~\bibnamefont
  {Stech}},\ }\href {\doibase 10.1103/PhysRevD.62.014006} {\bibfield  {journal}
  {\bibinfo  {journal} {Phys. Rev.}\ }\textbf {\bibinfo {volume} {D62}},\
  \bibinfo {pages} {014006} (\bibinfo {year} {2000})},\ \Eprint
  {http://arxiv.org/abs/hep-ph/0001113} {arXiv:hep-ph/0001113 [hep-ph]}
  \BibitemShut {NoStop}%
\bibitem [{\citenamefont {Bajc}\ \emph {et~al.}(1996)\citenamefont {Bajc},
  \citenamefont {Fajfer},\ and\ \citenamefont {Oakes}}]{Bajc:1995km}%
  \BibitemOpen
  \bibfield  {author} {\bibinfo {author} {\bibfnamefont {B.}~\bibnamefont
  {Bajc}}, \bibinfo {author} {\bibfnamefont {S.}~\bibnamefont {Fajfer}}, \ and\
  \bibinfo {author} {\bibfnamefont {R.~J.}\ \bibnamefont {Oakes}},\ }\href
  {\doibase 10.1103/PhysRevD.53.4957} {\bibfield  {journal} {\bibinfo
  {journal} {Phys. Rev.}\ }\textbf {\bibinfo {volume} {D53}},\ \bibinfo {pages}
  {4957} (\bibinfo {year} {1996})},\ \Eprint
  {http://arxiv.org/abs/hep-ph/9511455} {arXiv:hep-ph/9511455 [hep-ph]}
  \BibitemShut {NoStop}%
\bibitem [{\citenamefont {Khodjamirian}(2020)}]{Khodjamirian:2020btr}%
  \BibitemOpen
  \bibfield  {author} {\bibinfo {author} {\bibfnamefont {A.}~\bibnamefont
  {Khodjamirian}},\ }\href@noop {} {\emph {\bibinfo {title} {{Hadron Form
  Factors}: {From Basic Phenomenology to QCD Sum Rules}}}}\ (\bibinfo
  {publisher} {CRC Press, Taylor \& Francis Group},\ \bibinfo {address} {Boca
  Raton, FL, USA},\ \bibinfo {year} {2020})\BibitemShut {NoStop}%
\bibitem [{\citenamefont {Escribano}\ \emph {et~al.}(2014)\citenamefont
  {Escribano}, \citenamefont {Gonz\'alez-Sol\'\i{}s}, \citenamefont {Jamin},\
  and\ \citenamefont {Roig}}]{Escribano:2014joa}%
  \BibitemOpen
  \bibfield  {author} {\bibinfo {author} {\bibfnamefont {R.}~\bibnamefont
  {Escribano}}, \bibinfo {author} {\bibfnamefont {S.}~\bibnamefont
  {Gonz\'alez-Sol\'\i{}s}}, \bibinfo {author} {\bibfnamefont {M.}~\bibnamefont
  {Jamin}}, \ and\ \bibinfo {author} {\bibfnamefont {P.}~\bibnamefont {Roig}},\
  }\href {\doibase 10.1007/JHEP09(2014)042} {\bibfield  {journal} {\bibinfo
  {journal} {JHEP}\ }\textbf {\bibinfo {volume} {09}},\ \bibinfo {pages} {042}
  (\bibinfo {year} {2014})},\ \Eprint {http://arxiv.org/abs/1407.6590}
  {arXiv:1407.6590 [hep-ph]} \BibitemShut {NoStop}%
\bibitem [{\citenamefont {Jamin}\ \emph {et~al.}(2000)\citenamefont {Jamin},
  \citenamefont {Oller},\ and\ \citenamefont {Pich}}]{Jamin:2000wn}%
  \BibitemOpen
  \bibfield  {author} {\bibinfo {author} {\bibfnamefont {M.}~\bibnamefont
  {Jamin}}, \bibinfo {author} {\bibfnamefont {J.~A.}\ \bibnamefont {Oller}}, \
  and\ \bibinfo {author} {\bibfnamefont {A.}~\bibnamefont {Pich}},\ }\href
  {\doibase 10.1016/S0550-3213(00)00479-X} {\bibfield  {journal} {\bibinfo
  {journal} {Nucl. Phys. B}\ }\textbf {\bibinfo {volume} {587}},\ \bibinfo
  {pages} {331} (\bibinfo {year} {2000})},\ \Eprint
  {http://arxiv.org/abs/hep-ph/0006045} {arXiv:hep-ph/0006045} \BibitemShut
  {NoStop}%
\bibitem [{\citenamefont {Bethe}(1949)}]{Bethe:1949yr}%
  \BibitemOpen
  \bibfield  {author} {\bibinfo {author} {\bibfnamefont {H.~A.}\ \bibnamefont
  {Bethe}},\ }\href {\doibase 10.1103/PhysRev.76.38} {\bibfield  {journal}
  {\bibinfo  {journal} {Phys. Rev.}\ }\textbf {\bibinfo {volume} {76}},\
  \bibinfo {pages} {38} (\bibinfo {year} {1949})}\BibitemShut {NoStop}%
\bibitem [{\citenamefont {Albertus}\ \emph {et~al.}(2005)\citenamefont
  {Albertus}, \citenamefont {Hernandez}, \citenamefont {Nieves},\ and\
  \citenamefont {Verde-Velasco}}]{Albertus:2005vd}%
  \BibitemOpen
  \bibfield  {author} {\bibinfo {author} {\bibfnamefont {C.}~\bibnamefont
  {Albertus}}, \bibinfo {author} {\bibfnamefont {E.}~\bibnamefont {Hernandez}},
  \bibinfo {author} {\bibfnamefont {J.}~\bibnamefont {Nieves}}, \ and\ \bibinfo
  {author} {\bibfnamefont {J.~M.}\ \bibnamefont {Verde-Velasco}},\ }\href
  {\doibase 10.1103/PhysRevD.71.113006} {\bibfield  {journal} {\bibinfo
  {journal} {Phys. Rev. D}\ }\textbf {\bibinfo {volume} {71}},\ \bibinfo
  {pages} {113006} (\bibinfo {year} {2005})},\ \Eprint
  {http://arxiv.org/abs/hep-ph/0502219} {arXiv:hep-ph/0502219} \BibitemShut
  {NoStop}%
\bibitem [{\citenamefont {Lubicz}\ \emph {et~al.}(2017)\citenamefont {Lubicz},
  \citenamefont {Riggio}, \citenamefont {Salerno}, \citenamefont {Simula},\
  and\ \citenamefont {Tarantino}}]{Lubicz:2017syv}%
  \BibitemOpen
  \bibfield  {author} {\bibinfo {author} {\bibfnamefont {V.}~\bibnamefont
  {Lubicz}}, \bibinfo {author} {\bibfnamefont {L.}~\bibnamefont {Riggio}},
  \bibinfo {author} {\bibfnamefont {G.}~\bibnamefont {Salerno}}, \bibinfo
  {author} {\bibfnamefont {S.}~\bibnamefont {Simula}}, \ and\ \bibinfo {author}
  {\bibfnamefont {C.}~\bibnamefont {Tarantino}} (\bibinfo {collaboration}
  {ETM}),\ }\href {\doibase 10.1103/PhysRevD.96.054514,
  10.1103/PhysRevD.99.099902} {\bibfield  {journal} {\bibinfo  {journal} {Phys.
  Rev.}\ }\textbf {\bibinfo {volume} {D96}},\ \bibinfo {pages} {054514}
  (\bibinfo {year} {2017})},\ \bibinfo {note} {[Erratum: Phys. Rev. D99, no.9,
  099902 (2019)]},\ \Eprint {http://arxiv.org/abs/1706.03017} {arXiv:1706.03017
  [hep-lat]} \BibitemShut {NoStop}%
\bibitem [{\citenamefont {Aitala}\ \emph {et~al.}(2002)\citenamefont {Aitala}
  \emph {et~al.}}]{Aitala:2002kr}%
  \BibitemOpen
  \bibfield  {author} {\bibinfo {author} {\bibfnamefont {E.~M.}\ \bibnamefont
  {Aitala}} \emph {et~al.} (\bibinfo {collaboration} {E791}),\ }\href {\doibase
  10.1103/PhysRevLett.89.121801} {\bibfield  {journal} {\bibinfo  {journal}
  {Phys. Rev. Lett.}\ }\textbf {\bibinfo {volume} {89}},\ \bibinfo {pages}
  {121801} (\bibinfo {year} {2002})},\ \Eprint
  {http://arxiv.org/abs/hep-ex/0204018} {arXiv:hep-ex/0204018} \BibitemShut
  {NoStop}%
\bibitem [{\citenamefont {Bonvicini}\ \emph {et~al.}(2008)\citenamefont
  {Bonvicini} \emph {et~al.}}]{CLEO:2008jus}%
  \BibitemOpen
  \bibfield  {author} {\bibinfo {author} {\bibfnamefont {G.}~\bibnamefont
  {Bonvicini}} \emph {et~al.} (\bibinfo {collaboration} {CLEO}),\ }\href
  {\doibase 10.1103/PhysRevD.78.052001} {\bibfield  {journal} {\bibinfo
  {journal} {Phys. Rev. D}\ }\textbf {\bibinfo {volume} {78}},\ \bibinfo
  {pages} {052001} (\bibinfo {year} {2008})},\ \Eprint
  {http://arxiv.org/abs/0802.4214} {arXiv:0802.4214 [hep-ex]} \BibitemShut
  {NoStop}%
\bibitem [{\citenamefont {Ablikim}\ \emph {et~al.}(2019)\citenamefont {Ablikim}
  \emph {et~al.}}]{Ablikim:2018upe}%
  \BibitemOpen
  \bibfield  {author} {\bibinfo {author} {\bibfnamefont {M.}~\bibnamefont
  {Ablikim}} \emph {et~al.} (\bibinfo {collaboration} {BESIII}),\ }\href
  {\doibase 10.1103/PhysRevLett.122.061801} {\bibfield  {journal} {\bibinfo
  {journal} {Phys. Rev. Lett.}\ }\textbf {\bibinfo {volume} {122}},\ \bibinfo
  {pages} {061801} (\bibinfo {year} {2019})},\ \Eprint
  {http://arxiv.org/abs/1811.02911} {arXiv:1811.02911 [hep-ex]} \BibitemShut
  {NoStop}%
\bibitem [{\citenamefont {del Amo~Sanchez}\ \emph {et~al.}(2011)\citenamefont
  {del Amo~Sanchez} \emph {et~al.}}]{BaBar:2010wqe}%
  \BibitemOpen
  \bibfield  {author} {\bibinfo {author} {\bibfnamefont {P.}~\bibnamefont {del
  Amo~Sanchez}} \emph {et~al.} (\bibinfo {collaboration} {BaBar}),\ }\href
  {\doibase 10.1103/PhysRevD.83.052001} {\bibfield  {journal} {\bibinfo
  {journal} {Phys. Rev. D}\ }\textbf {\bibinfo {volume} {83}},\ \bibinfo
  {pages} {052001} (\bibinfo {year} {2011})},\ \Eprint
  {http://arxiv.org/abs/1011.4190} {arXiv:1011.4190 [hep-ex]} \BibitemShut
  {NoStop}%
\bibitem [{\citenamefont {Aaij}\ \emph {et~al.}(2019)\citenamefont {Aaij} \emph
  {et~al.}}]{Aaij:2018hik}%
  \BibitemOpen
  \bibfield  {author} {\bibinfo {author} {\bibfnamefont {R.}~\bibnamefont
  {Aaij}} \emph {et~al.} (\bibinfo {collaboration} {LHCb}),\ }\href {\doibase
  10.1007/JHEP03(2019)176} {\bibfield  {journal} {\bibinfo  {journal} {JHEP}\
  }\textbf {\bibinfo {volume} {03}},\ \bibinfo {pages} {176} (\bibinfo {year}
  {2019})},\ \Eprint {http://arxiv.org/abs/1810.03138} {arXiv:1810.03138
  [hep-ex]} \BibitemShut {NoStop}%
\bibitem [{\citenamefont {Kampf}\ \emph {et~al.}(2018)\citenamefont {Kampf},
  \citenamefont {Novotn{\'y}},\ and\ \citenamefont
  {Sanchez-Puertas}}]{Kampf:2018wau}%
  \BibitemOpen
  \bibfield  {author} {\bibinfo {author} {\bibfnamefont {K.}~\bibnamefont
  {Kampf}}, \bibinfo {author} {\bibfnamefont {J.}~\bibnamefont {Novotn{\'y}}},
  \ and\ \bibinfo {author} {\bibfnamefont {P.}~\bibnamefont
  {Sanchez-Puertas}},\ }\href {\doibase 10.1103/PhysRevD.97.056010} {\bibfield
  {journal} {\bibinfo  {journal} {Phys. Rev.}\ }\textbf {\bibinfo {volume}
  {D97}},\ \bibinfo {pages} {056010} (\bibinfo {year} {2018})},\ \Eprint
  {http://arxiv.org/abs/1801.06067} {arXiv:1801.06067 [hep-ph]} \BibitemShut
  {NoStop}%
\end{thebibliography}%

\end{document}